\documentclass[journal,11pt,onecolumn]{IEEEtran}
\usepackage{amsmath,amssymb,amsfonts}
\usepackage{algorithmic,algorithm}
\usepackage[caption=false,font=normalsize,labelfont=sf,textfont=sf]{subfig}
\usepackage{graphicx}
\usepackage{float}
\usepackage{url}
\usepackage{cite}
\usepackage{cleveref}
\ifCLASSOPTIONonecolumn
\usepackage{setspace}
\fi
\allowdisplaybreaks

\def\E{\mathbb{E}}
\def\Var{\mathbb{V}}

\begin{document}

\title{Gaussian Kernel Variance For an Adaptive Learning Method on Signals Over Graphs}
\author{Yue~Zhao,~\IEEEmembership{Student Member,~IEEE,}~and~Ender~Ayanoglu,~\IEEEmembership{Fellow,~IEEE}
\thanks{The authors are with the Center for Pervasive Computing and Communications, Department of Electrical Engineering and Computer Science, University of California Irvine, Irvine, CA, USA.}
\thanks{Manuscript received December 6, 2021; revised March 5, 2022; accepted April 14, 2022.}
}
\maketitle
\begin{abstract}
This paper discusses a special kind of a simple yet possibly powerful algorithm, called single-kernel Gradraker (SKG), which is an adaptive learning method predicting unknown nodal values in a network using known nodal values and the network structure. We aim to find out how to configure the special kind of the model in applying the algorithm. To be more specific, we focus on SKG with a Gaussian kernel and specify how to find a suitable variance for the kernel. To do so, we introduce two variables with which we are able to set up requirements on the variance of the Gaussian kernel to achieve (near-) optimal performance and can better understand how SKG works. Our contribution is that we introduce two variables as analysis tools, illustrate how predictions will be affected under different Gaussian kernels, and provide an algorithm finding a suitable Gaussian kernel for SKG with knowledge about the training network. Simulation results on real datasets are provided.
\end{abstract}
\section{Introduction}
\label{sec:intro}

\IEEEPARstart{C}{omplex} systems can be described by means of graphs. An example is abstracting the citing behavior of a set of papers into a two-dimensional matrix \cite{CoraDataset}. In those abstracted graphs, links denote relationships between nodes. Nodes can carry information only about themselves, which are called nodal values. Nodal values can be inaccessible for part of the nodes. They might be, however, inferred from the nodal values of other nodes which are known, together with the network structure. 

The problem of inferring unknown nodal values can be considered as semi-supervised learning \cite{SemiLearn}. The estimation problem can also be dealt with by techniques of signal reconstruction \cite{GraphSignalSampRecon,GraphSignalRecon} or signal interpolation \cite{GraphSignalIntepol} in the emerging field of Graph Signal Processing (GSP) \cite{DSPOnG,DSPOnGFrequencyAnalysis,GSPEmergingField,GSP2018}. Noticing that the majority of GSP studies focus on linear graph filters \cite{LinGraphFilter}, researchers try to design nonlinear graph filters \cite{NonlinGraphFilter} to cope with the nonlinearity residing in signals over graphs. Another way to consider the nonlinearity is to apply kernel methods on graphs \cite{GraphKernel2002,GraphKernel2003,GraphKernel2017}. Nonlinear algorithms usually outperform their linear counterparts, however, their computational costs also grow faster with more known nodes, making them less practical for large networks.

Fortunately, \cite{RandomFeatures} discovers that for shift-invariant kernels, the value can be approximated using random features drawn from the Fourier transform of the kernel. Based on that, \cite{OnlineGradraker} proposes a simple yet possibly powerful algorithm called \textbf{gr}aph-\textbf{ad}aptive learning method using \textbf{ra}ndom feature approximation with multiple \textbf{ker}nels, abbreviated as Gradraker. The algorithm takes connection information of a node as input and trains a model to output the corresponding nodal value under supervised learning. Since only vector additions and multiplications are needed, acquiring predictions is convenient and updating the model parameters online becomes possible, which makes the Gradraker algorithm promising to large dynamic networks. What is more, the usage of different kernels or their mixtures might also extend usable applications. Additionally, the Gradraker algorithm reserves nodal privacy to some extent thanks to the incorporation of the random feature approximation \cite{RandomFeatures}. So, we think the algorithm or its variants are applicable for an extensive set of scenarios, e.g., traffic dynamic estimation, account anomaly detection in social software, recommendation systems, etc. The authors of \cite{OnlineGradraker} have shown the impressive performance of the algorithm in terms of Normalized Mean Square Error (NMSE) and its low complexity. Authors of \cite{GradrakerMultiHop} propose a similar algorithm, Graph Kernel Least Mean Squares-Random Fourier Features (GKLMS-RFF), which contains the same model but takes graph-filtered nodal value time sequence of a node as input, instead of the adjacency vector of a node, and provide the convergence condition. 
Gradraker is extended to exploit multi-hop information for estimation in \cite{GradrakerMultiHop}, and has the potential to be applied on multi-layer graphs.

There are few papers guiding how to configure the model in Gradraker-like algorithms, especially in a theoretical view. We aim to fill this gap. The purpose of doing so is not only to have guidance in configuration, but also to have a better understanding on the pros and cons of the algorithm, recognizing its applicable situations and possibly giving hints on the design of its variants. We choose the Single-kernel Gradraker (SKG) algorithm as the entry point. A Gradraker model consists of several SKG models which share the input of the Gradraker model. Each component SKG model outputs an estimation which is then used for acquiring the final estimation of the Gradraker model via an aggregation algorithm. That is, SKG models are building blocks of Gradraker-based algorithms and Gradraker performance is highly influenced by the best performance among all the components. Thus, understanding SKG performance in a detailed manner is of great importance for future studies.

To achieve the best performance for an SKG model, there are a few hyperparameters, i.e., the loss function, the learning rate, the number of repeated times for training, the number of random features, and the Gaussian kernel variance, which should be properly chosen. The loss function is selected based on applications. For instance, Least Squares (LS) loss function is usually applied in regression problems. A suitable value of the learning rate is proposed in \cite{LearningRate}. The number of repeated times for training can be found by techniques like monitoring validation loss during training and stopping training when the validation loss does not improve \cite{callback} as done in the field of machine learning. The number of random features does not play a major role affecting the model performance once it is sufficiently large. Thus, we will focus on the problem of choosing a suitable kernel for a training set for there is no discussion on it prior to our paper to the best of our knowledge. The study of choosing a suitable kernel is not trivial. Gradraker is proposed using a kernel dictionary with multiple kernels, letting the algorithm choose suitable ones. However, noticing that the computational cost grows linearly with the size of the dictionary, when inappropriate kernels are inside, computation complexity grows beyond what is needed with appropriate kernels. What is worse, if the kernel dictionary does not contain any suitable kernel, the performance would be bad, and adding more kernels blindly to the dictionary may not be beneficial. So, distinguishing suitable kernels helps to achieve the best performance with the lowest cost. Among many kinds of shift-invariant kernels \cite{RandomFeatures}, e.g., Gaussian kernels, Laplacian kernels, and Cauchy kernels, we will focus on Gaussian kernels. Noting that the kernel being used models how similarity changes with difference, and that the laws of large numbers indicate wide application of the Gaussian distribution, it is intuitive to use Gaussian kernels in most situations \cite{OnlineGradraker,GradrakerTimeInfo,GradrakerMultiHop}. For this reason, in this paper we will discuss SKG with a Gaussian kernel in detail.

To do an analysis on SKG, we build an analysis framework based on two new variables, the similarity measure and the contribution weight. Our contributions are
\begin{itemize}
    \item Introducing two variables to do analysis regarding the SKG algorithm;
    \item Illustrating the impact of different Gaussian kernel variance on prediction performance;
    \item Providing an algorithm to find a suitable Gaussian variance for an SKG model given a training dataset.
\end{itemize}

The rest of the paper is organized as follows. Section \ref{sec:steps} gives a review of SKG. Mathematical tools will be introduced in Section \ref{sec:math} followed by the impact analysis of different Gaussian kernels and an algorithm to find a suitable Gaussian kernel for a given training set in Section \ref{sec:how}. Section \ref{sec:sim} shows simulation results to verify properties of introduced variables and effectiveness of the proposed algorithm. Discussions and conclusions are provided in Section \ref{sec:disc} and Section \ref{sec:conclu}, respectively.

\textit{Notation}: Vectors are denoted by bold lowercase characters. Matrices are denoted by bold uppercase characters. The symbol $(\cdot)^\top$ denotes the transpose of a vector or a matrix. The $(m,n)$-th ($m$-th) element in an $M\times N$ matrix $\textbf{A}$ ($M\times1$ column vector or $1\times M$ row vector $\textbf{a}$) is denoted by $[\textbf{A}]_{m,n}$ ($[\textbf{a}]_{m}$) where $1\leq m\leq M$ and $1\leq n\leq N$. The $l_2$ norm is denoted by $\|\cdot\Vert$. The notation $|\cdot\vert$ represents the absolute value for a number, or the cardinality of a set. The (conditional) expectation is denoted by $\E[\cdot]$ ($\E[\cdot|\text{given variable}]$), and the (conditional) variance is denoted by $\Var[\cdot]$ (($\Var[\cdot|\text{given variable}]$)). The minimum value of a sequence $\textbf{s}$ is denoted by $\min\textbf{s}$.

\section{Prior Work: Steps When Applying SKG}
\label{sec:steps}

The basic sequence of steps applying SKG is preparing the training set, building up and initializing the model, training sequentially, and performing prediction (updating trainable parameters if available).

\subsection{Preparing a Training Set}
Let there be a set of sampled nodes $\mathcal{V}=\{v_n\}_{n=1}^N$ with known nodal values $\{y_n\}_{n=1}^N$, a set of referencing nodes $\mathcal{V}_r=\{v_{r,m}\}_{m=1}^M$, and a description of connection between the two sets of nodes. Note that nodal values of referenced nodes do not play a role. The description of connection can be in the form of a matrix $\textbf{A}$, of size $M\times N$, whose element $[\textbf{A}]_{m,n}$ is $0$ if the sampled node $v_n$ is not connected with the referencing node $v_{r,m}$, or $1$ if the two nodes are connected in the case of unweighted graphs, or the weight over the edge connecting the two nodes in the case of weighted graphs. Note that a column of $\textbf{A}$ reports the description of connection between the corresponding sampled node and all the referencing nodes. So, we call the description vector \textit{adjacency vector} of the sampled node. Denote the adjacency vector of the sampled node $v_n$ as $\textbf{a}_n$ with size $M\times 1$. Combining with the nodal value $y_n$ of $v_n$, we get the pair $(\textbf{a}_n,y_n)$ for $v_n$, and the set of pairs $\{(\textbf{a}_n,y_n)\}_{n=1}^N$ is called the training set.

The sampled node set $\mathcal{V}$ and the referencing node set $\mathcal{V}_r$ are not necessarily the same. In \cite{OnlineGradraker}, $\mathcal{V}=\mathcal{V}_r$, and thus the $N$ adjacency vectors can be formatted in an adjacency matrix of size $N\times N$. In our paper, we generalize applicable scenarios such that $\mathcal{V}_r$ can be any set of nodes, without the need to modify the Gradraker algorithm. 

\subsection{Building Up the Model and Initialization}
The model takes an adjacency vector $\textbf{a}_n$ of $v_n\in\mathcal{V}$ as input. The first part of the SKG model is for acquiring a nonlinear transform $z(\textbf{a}_n)$ of the input $\textbf{a}_n$ through a nonlinear mapping $z:\mathbb{R}^{M\times1}\mapsto\mathbb{R}^{2D\times1}$. Specifically,
\ifCLASSOPTIONonecolumn
\begin{equation}
\label{equ:nonlinear}
z(\textbf{a}_n)=[\sin(\boldsymbol{\xi}_1^\top\textbf{a}_n),\sin(\boldsymbol{\xi}_2^\top\textbf{a}_n),...,\sin(\boldsymbol{\xi}_D^\top\textbf{a}_n),
\cos(\boldsymbol{\xi}_1^\top\textbf{a}_n),\cos(\boldsymbol{\xi}_2^\top\textbf{a}_n),...,\cos(\boldsymbol{\xi}_D^\top\textbf{a}_n)]^\top/\sqrt{D}
\end{equation}
\else
\begin{multline}
\label{equ:nonlinear}
z(\textbf{a}_n)=[\sin(\boldsymbol{\xi}_1^\top\textbf{a}_n),\sin(\boldsymbol{\xi}_2^\top\textbf{a}_n),...,\sin(\boldsymbol{\xi}_D^\top\textbf{a}_n),\\
\cos(\boldsymbol{\xi}_1^\top\textbf{a}_n),\cos(\boldsymbol{\xi}_2^\top\textbf{a}_n),...,\cos(\boldsymbol{\xi}_D^\top\textbf{a}_n)]^\top/\sqrt{D}
\end{multline}
\fi
where $\{\boldsymbol{\xi}_i\}_{i=1}^D$ are random features \cite{RandomFeatures} drawn from a distribution which is the Fourier transform of the kernel $\kappa$ in SKG. Note we have to manually choose $\kappa$. Recall that we will focus on Gaussian kernels in the paper, then the problem is reduced to choosing a variance $\sigma^2$ for the Gaussian kernel. Once the kernel is chosen, the following claim is helpful to generate $\{\boldsymbol{\xi}_i\}_{i=1}^D$.

\textbf{Claim 1}. Supposing a Gaussian kernel $\kappa$ with variance of $\sigma^2$, i.e., $\kappa(\textbf{x}_1,\textbf{x}_2)=e^{-\frac{\|\textbf{x}_1-\textbf{x}_2\Vert^2}{2\sigma^2}}$, random features $\{\boldsymbol{\xi}_i\}_{i=1}^D$ should be drawn from the Gaussian distribution $\mathcal{N}(0,\sigma^{-2}\textbf{I})$ when using the random feature approximation for $\kappa$ \cite{RandomFeatures,OnlineGradraker}.

\textit{Proof}. The proofs of all the claims in the paper are provided in the Appendix. \\

According to \textrm{\textbf{Claim 1}}, random features $\{\boldsymbol{\xi}_i\}_{i=1}^D$ should follow $\mathcal{N}(0,\sigma^{-2}\textbf{I})$. Note that $D$ is also preselected and that random features are fixed during training and predicting phases once they are chosen. 

Then, $z(\textbf{a}_n)$ goes through the second part of the model, which is linear, and provides a prediction as
\begin{equation}
\label{equ:predict}
\hat{f}_n=\boldsymbol{\theta}^\top z(\textbf{a}_n)
\end{equation}
where $\hat{f}_n$ denotes the prediction. The column vector $\boldsymbol{\theta}$ whose size is $2D\times 1$ is the trainable parameter.

Prior to the training phase, the trainable parameter is initialized as $\boldsymbol{\theta}_0=\textbf{0}$. Since $\boldsymbol{\theta}$ is updated every time, we will use $\boldsymbol{\theta}_t$ to denote the $\boldsymbol{\theta}$ value at the end of time (iteration) $t$.


\subsection{Sequential Training}
The parameter $\boldsymbol{\theta}$ is updated by the gradient descent algorithm, i.e., $\boldsymbol{\theta}_t,1\leq t\leq T$ where $T$ denotes the training duration is updated via
\begin{equation}
\label{equ:thetaupdate}
\boldsymbol{\theta}_t=\boldsymbol{\theta}_{t-1}-\eta\nabla_{\boldsymbol{\theta}}\mathcal{L}_t
\end{equation}
where $\nabla_{\boldsymbol{\theta}}\mathcal{L}_t$ is the gradient of the loss function $\mathcal{L}$ with respect to $\boldsymbol{\theta}$ at time $t$ and $\eta$ is the preselected learning rate. Noting that LS loss is used, we have $\mathcal{L}(y_{true},\hat{f})=(y_{true}-\hat{f})^2$. Then, the gradient at time $t$ which is employed in (\ref{equ:thetaupdate}) is
\begin{equation}
\label{equ:gradient}
\nabla_{\boldsymbol{\theta}}\mathcal{L}_t=-2(y_{n_t}-\hat{f}_{n_t})z(\textbf{a}_{n_t})
\end{equation}
where $y_{n_t}$, $\hat{f}_{n_t}$, and $\textbf{a}_{n_t}$ are the true nodal value, the prediction, and the adjacency vector of the node used at time $t$, respectively. Note that $\textbf{a}_{n_t}$ is not any specific adjacency vector but random because there is no assumed order for sampled nodes being processed. For notational simplicity, we will use $\textbf{a}_t$, $\hat{f}_t$, and $y_t$ instead of $\textbf{a}_{n_t}$, $\hat{f}_{n_t}$, and $y_{n_t}$ from now on. 

In \cite{OnlineGradraker} and other papers about kernel-based predicting methods, sometimes an overfitting-controlling term is summed with the squared difference $(y_{pred}-y_{true})^2$ in calculating the loss $\mathcal{L}(y_{pred},y_{true})$. However, the overfitting-controlling term does not greatly affect the level of best performance of SKG achieved on a graph signal, so we ignore it in our analysis.

During the training phase, we process a sampled node at a time, i.e., getting its adjacency vector which is then put through the model to get a prediction, and updating the model. The process is repeated until all sampled nodes are processed. It is possible for the training set to be used multiple times during the training phase, and the number of times the training set is repeatedly used is called the number of epochs, denoted by $E$. The parameter $E$ is also preselected. The training duration is $T=EN$.

\subsection{Predicting}
When the adjacency vector $\textbf{a}'$ of a tested node $v'$ to the set of referencing nodes is known, we can use the well-trained model to predict the nodal value. To that end, we first acquire the nonlinear transform $z(\textbf{a}')$ via (\ref{equ:nonlinear}), and then get a prediction via (\ref{equ:predict}). If its true value can be known, the trainable parameter $\boldsymbol{\theta}$ of the model can be updated via (\ref{equ:thetaupdate}).

\section{Mathematical Tools}
\label{sec:math}
A convincing way to illustrate the influence of $\sigma^2$ on predictions is to express predictions explicitly in terms of $\sigma^2$. It is a hard problem due to the training process. Alternatively, we express predictions as weighted averages of observations, where the influence of $\sigma^2$ on the weights is easier to show. To do so, we have to introduce two variables. One is used as weights of observations in predictions, called \textit{contribution weight}. The other one, called \textit{similarity measure}, is an intermediate variable in finding contribution weights. In this section, we give definitions of the two variables, and state their properties; preparing for analysis on how prediction behaves under different $\sigma^2$ in the next section. We introduce the \textit{similarity measure} $B_{i,j}$ first as it is basic to the definition of the \textit{contribution weight} $F_{i,j}$.

\subsection{Similarity Measure}
\label{subsec:bij}
The definition of the similarity measure $B_{i,j}$ for the pair of nodes seen at time $i$ and $j$, $1\leq i<j\leq T+1$, is
\begin{equation}
\label{equ:bijdef}
B_{i,j}\triangleq2\eta z^\top(\textbf{a}_i)z(\textbf{a}_j)
\end{equation}
where $\textbf{a}_i$ and $\textbf{a}_j$ are the adjacency vectors of nodes used at time $i$ and time $j$. As its name suggests, $B_{i,j}$ can be seen as a similarity measure between $\textbf{a}_i$ and $\textbf{a}_j$ recalling that $\kappa(\textbf{a}_i,\textbf{a}_j)\approx z^\top(\textbf{a}_i)z(\textbf{a}_j)$ \cite{OnlineGradraker}, and that a kernel function is a form of similarity measure.

Note that $B_{i,j}$ is a random number. Its randomness comes from both the random features $\{\boldsymbol{\xi}_i\}_{i=1}^D$ and $\textbf{a}_i$ and $\textbf{a}_j$ because adjacency vectors vary among different datasets. Even for a given dataset, $\textbf{a}_i$ and $\textbf{a}_j$ cannot be determined because of the random sampling for the sampled nodes. Considering the uniform distribution of when a specific node is processed within an epoch, the distribution of $B_{i,j}$ is identical for all qualified pairs of $i$ and $j$. We keep the indices to denote the time when the adjacency vectors are processed.

Since weights of observations in predictions build on similarity measures $B_{i,j}$, studying properties of $B_{i,j}$ not only helps understanding how $\sigma^2$ changes $B_{i,j}$, but also paves a path to the impact of $\sigma^2$ on observations weights. So, we illustrate two properties of $B_{i,j}$, \textit{exponential approximation} and \textit{positive average}, in the following.

Substituting (\ref{equ:nonlinear}) into (\ref{equ:bijdef}), we have
\begin{align}
\label{equ:bijval}
B_{i,j}=&\frac{2\eta}{D}\sum_{k=1}^D[\sin(\boldsymbol{\xi}_k^\top\textbf{a}_i)\sin(\boldsymbol{\xi}_k^\top\textbf{a}_j)+\cos(\boldsymbol{\xi}_k^\top\textbf{a}_i)\cos(\boldsymbol{\xi}_k^\top\textbf{a}_j)]\nonumber\\
=&2\eta\frac{\sum_{k=1}^D\cos[\boldsymbol{\xi}_k^\top(\textbf{a}_i-\textbf{a}_j)]}{D}\nonumber\\
=&2\eta\frac{\sum_{k=1}^D\cos[\boldsymbol{\xi}_k^\top\textbf{d}_{i,j}]}{D}
\end{align}
where $\textbf{d}_{i,j}=\textbf{a}_i-\textbf{a}_j$. It is shown in (\ref{equ:bijval}) that $B_{i,j}$ is actually the sample average of $D$ terms of $\cos(C_{k,i,j})$ where $C_{k,i,j}=\boldsymbol{\xi}_k^\top\textbf{d}_{i,j}$ multiplied by a scalar $2\eta$. Whereas, $\textbf{d}_{i,j}$ is a random vector with respect to different training data and the random processing order of sampled nodes, $\{\boldsymbol{\xi}_k\}_{k=1}^D$ are related to model configuration. We will focus on how $B_{i,j}$ changes with respect to $\{\boldsymbol{\xi}_k\}_{k=1}^D$ (viewing $\textbf{a}_i$ and $\textbf{a}_j$ as given for now). Recalling from \textrm{\textbf{Claim 1}} that elements of $\boldsymbol{\xi}_k,k=1,...,D$ are independently and identically distributed (i.i.d.) Gaussian random numbers with variance $\sigma^{-2}$, we know that $C_{k,i,j}$ follows $\mathcal{N}(0,\|\textbf{d}_{i,j}\Vert^2/\sigma^2)$ for a given $\textbf{d}_{i,j}$. Notice $\{C_{k,i,j}\}_{k=1}^D$ are i.i.d. because of i.i.d. $\{\boldsymbol{\xi}_k\}_{k=1}^D$. Supposing $D$ is sufficiently large, we can follow the weak law of large numbers and get
\begin{equation}
\label{equ:bijexpect}
    B_{i,j}\cong2\eta\E[\cos(C_{k,i,j})|\textbf{d}_{i,j}].
\end{equation}
Note $B_{i,j}$ is a random number but varies in a small range given $\textbf{d}_{i,j}$ and a sufficiently large $D$. The following claim can be useful to get the explicit expression of the conditional expectation.

\textbf{Claim 2}. Suppose $X$ is a Gaussian random number such that $X\thicksim\mathcal{N}(0,\sigma_X^2)$. Then, 
\begin{equation}
    \label{equ:expect}
    \E[\cos(X)]=e^{-\frac{\sigma_X^2}{2}},
\end{equation}
\begin{equation}
    \label{equ:var}
    \Var[\cos(X)]=\frac{1}{2}(e^{-\sigma_X^2}-1)^2.
\end{equation}

\subsubsection{Exponential Approximation}
Substituting (\ref{equ:expect}) into (\ref{equ:bijexpect}), we get the exponential approximation
\begin{equation}
\label{equ:eb}
B_{i,j}\cong2\eta\E[\cos(C_{k,i,j})|\textbf{d}_{i,j}]=2\eta e^{-\frac{\|\textbf{d}_{i,j}\Vert^2}{2\sigma^2}}
\end{equation}
when $D$ is sufficiently large.

The equality in (\ref{equ:eb}) is the same expression as the random feature approximation \cite{RandomFeatures,OnlineGradraker}, but in a reverse order. Recall that for a Gaussian kernel $\kappa(\textbf{a}_i,\textbf{a}_j)=e^{-\frac{\|\textbf{a}_i-\textbf{a}_j\Vert^2}{2\sigma^2}}$, its mathematical expression of the random feature approximation is $\kappa(\textbf{a}_i,\textbf{a}_j)=e^{-\frac{\|\textbf{a}_i-\textbf{a}_j\Vert^2}{2\sigma^2}}\approx z^\top(\textbf{a}_i)z(\textbf{a}_j)$ where $z(\cdot)$ is defined in (\ref{equ:nonlinear}). If we replace $B_{i,j}$ and $\textbf{d}_{i,j}$ in (\ref{equ:eb}) with the definition in (\ref{equ:bijdef}) and $\textbf{a}_i-\textbf{a}_j$, respectively, we get $2\eta z^\top(\textbf{a}_i)z(\textbf{a}_j)\cong2\eta e^{-\frac{\|\textbf{a}_i-\textbf{a}_j\Vert^2}{2\sigma^2}}$ which is the same as the random feature approximation. The equality in (\ref{equ:eb}) explicitly shows how $\sigma^2$ affects $B_{i,j}$.

\subsubsection{Positive Average}
Taking expectation for (\ref{equ:bijval}), we get
\begin{equation}
\label{equ:ebij}
    \E\left[B_{i,j}\right]=2\eta\E\left[\E\left[\cos\left(C_{k,i,j}\right)|\textbf{d}_{i,j}\right]\right].
\end{equation}
In (\ref{equ:ebij}), $\E\left[B_{i,j}\right]$ is with respect to the
joint distribution of $\{\boldsymbol{\xi}_k\}_{k=1}^D$ and
$\textbf{d}_{i,j}$. On the right hand side, the inner expectation is
with respect to the conditional distribution of
$\{\boldsymbol{\xi}_k\}_{k=1}^D$ given
$\textbf{d}_{i,j}$ while the outer expectation is with respect to the
distribution of $\textbf{d}_{i,j}$.
Recall that when $\textbf{d}_{i,j}$ is given and $D$ is sufficiently large, $B_{i,j}$ varies within a small range and can be approximated via (\ref{equ:eb}). For a real dataset, $\textbf{d}_{i,j}$ is usually not deterministic but has a distribution under random sampling without replacement, and thus $B_{i,j}$ may greatly vary with different $\textbf{d}_{i,j}$ values. From \textrm{\textbf{Claim 2}}, it is known that $0<\E\left[\cos\left(C_{k,i,j}\right)|\textbf{d}_{i,j}\right]\leq 1$ for any $\textbf{d}_{i,j}$, where the equality holds when $\|\textbf{d}_{i,j}\Vert^2=0$. So, for real datasets where nonzero $\textbf{d}_{i,j}$ exists, we get from (\ref{equ:ebij}) that $0<\E\left[B_{i,j}\right]<2\eta$. 

\subsection{Contribution Weight}
As we mentioned before, we aim to express a prediction of SKG in terms of observations. We are able to do so via contribution weights introduced in the following. The definition of the contribution weights $F_{i,j}$ for the pair of nodes seen at time $i$ and $j$, $1\leq i<j\leq T+1$ is
\begin{equation}
\label{equ:fijdef}
F_{i,j}\triangleq
	\begin{cases}
	B_{i,j}, & \text{for } i=j-1,\\
	B_{i,j}-\sum_{k=i+1}^{j-1}B_{i,k}F_{k,j}, & \text{for } 1\leq i<j-1,
	\end{cases}
\end{equation}
and undefined otherwise. Because of the randomness in $B_{p,q},i\leq p\leq q\leq j$, $F_{i,j}$ is also a random number. The definition in (\ref{equ:fijdef}) indicates that $F_{i,j}$ is affected by $\sigma^2$ indirectly via $B_{i,j}$.

We show two useful properties for $F_{i,j}$, \textit{weighting} and \textit{conformity with $B_{i,j}$}.

\subsubsection{Weighting Property}
Firstly, the following claim shows $\{F_{i,j}\}_{i=1}^{j-1}$ are used as coefficients of previously seen nodal values in prediction.

\textbf{Claim 3}. Assume we are applying SKG on a training set. During the training phase, at time $t,1< t\leq T$, we have
\begin{equation}
\label{equ:hatft}
\hat{f}_t=\sum_{i=1}^{t-1}y_{i}F_{i,t}
\end{equation}
where $\hat{f}_t$ denotes the prediction at time $t$, $y_i$ denotes the true nodal value at training time $i$, and $\{F_{i,t}\}_{i=1}^{t-1}$ are defined as in (\ref{equ:fijdef}).\\

Although \textrm{\textbf{Claim 3}} mentions the training phase only, it is easy to extend (\ref{equ:hatft}) to the predicting (testing) phase. If the nodal value for a tested node is known somehow, the node will work as a sampled node and the SKG model can be trained further in which case (\ref{equ:hatft}) works fine for the tested node directly. If a tested node comes without a true nodal value, it will have no impact on the model, and all such nodes share exactly the same model in which case all these nodes can be seen as the node at time $T+1$. In practice, tested nodes with and without known nodal values may be mixed, however, only those with known nodal values will affect the model and later predictions. So, without loss of generality, we will only consider the case where nodal values for tested nodes are unknown. Then, the prediction for a tested node can be expressed as
\begin{equation}
\label{equ:hatfT}
\hat{f}_{T+1}=\sum_{i=1}^Ty_{i}F_{i,T+1}
\end{equation}
where $\hat{f}_{T+1}$ represents the prediction of the tested node. Note that at the time, we only show a prediction is a weighted summation of observations. Based on the following claim, we could step further and show a prediction can be a weighted average of observations using $F_{i,T+1}$.

\textbf{Claim 4}. According to the definition in (\ref{equ:fijdef}), we can get the expectation of the sum of $F_{i,T+1}$ with all qualified $i$, i.e., $1\leq i\leq T$, as
\begin{equation*}
\E\left[\sum_{i=1}^TF_{i,T+1}\right]=1-(1-b)^T
\end{equation*}
where $b=\E[B_{i,j}]$ and $1\leq i< j\leq T+1$.\\

It has been confirmed by the positive average property of $B_{i,j}$ in Section \ref{subsec:bij} that $0<b<2\eta$. For most cases where $\eta\ll1$, we get
\begin{equation}
\label{equ:dijsum}
	\lim_{T \to \infty}\E\left[\sum_{i=1}^TF_{i,T+1}\right]=1.
\end{equation}

In practice, the training duration $T$ is usually more than hundreds which is sufficient to get $\E\left[\sum_{i=1}^TF_{i,T+1}\right]\approx1$. The small variance of the summation is observed from experiments such that the summation is close to 1. Thus, using (\ref{equ:dijsum}) together with \textrm{\textbf{Claim 3}}, we can draw a conclusion that, when $T$ is suitable, the SKG prediction for a tested node is actually a weighted average of all previously seen nodal values, and how much contribution that the nodal value seen at time $i$ makes to the prediction is determined by the Contribution Weight $F_{i,T+1}$. We call this the weighting property of $F_{i,T+1}$.

\subsubsection{Conformity Between $B_{i,j}$ and $F_{i,j}$}
It can be observed from simulations that although $F_{i,j}$ for any $1<j\leq T+1$ increases about exponentially when $1\leq i<j$, $F_{i,j}$ tends to be greater when $B_{i,j}$ is obviously larger than $\E\left[B_{i,j}\right]$. Due to the recursive definition of $F_{i,j}$, the direct derivation between $B_{i,j}$ and $F_{i,j}$ becomes complex. So, we explain the conformity between $B_{i,j}$ and $F_{i,j}$ using induction.

For $j>1$, because $F_{j-1,j}=B_{j-1,j}$, there is no question that $B_{j-1,j}$ and $F_{j-1,j}$ will be both large or small. 

For $j>2$, we see
\ifCLASSOPTIONonecolumn
\begin{align*}
F_{j-2,j}&=B_{j-2,j}-B_{j-2,j-1}F_{j-1,j}=B_{j-2,j}-B_{j-2,j-1}B_{j-1,j}\\
&=B_{j-2,j}\bigg(1-\frac{B_{j-2.j-1}B_{j-1,j}}{B_{j-2,j}}\bigg).
\end{align*}
\else
\begin{align*}
F_{j-2,j}&=B_{j-2,j-1}-B_{j-2,j-1}F_{j-1,j}\\
&=B_{j-2,j}-B_{j-2,j-1}B_{j-1,j}\\
&=B_{j-2,j}\bigg(1-\frac{B_{j-2,j-1}B_{j-1,j}}{B_{j-2,j}}\bigg).
\end{align*}
\fi
Assuming a sufficiently large $D$,  we consider the exponential approximation of $B_{i,j}$ and get
\begin{align*}
\frac{B_{j-2,j-1}B_{j-1,j}}{B_{j-2,j}}&\approx\frac{2\eta e^{-\frac{\|\textbf{d}_{j-2,j-1}\Vert^2}{2\sigma^2}}\cdot2\eta e^{-\frac{\|\textbf{d}_{j-1,j}\Vert^2}{2\sigma^2}}}{2\eta e^{-\frac{\|\textbf{d}_{j-2,j}\Vert^2}{2\sigma^2}}}\\
&=2\eta e^{-\frac{\|\textbf{d}_{j-2,j-1}\Vert^2+\|\textbf{d}_{j-1,j}\Vert^2-\|\textbf{d}_{j-2,j}\Vert^2}{2\sigma^2}}.
\end{align*}
From Triangle Inequality, we know that $\|\textbf{d}_{j-2,j-1}\Vert^2+\|\textbf{d}_{j-1,j}\Vert^2-\|\textbf{d}_{j-2,j}\Vert^2\geq0$, so
\begin{equation*}
    0<\frac{B_{j-2,j-1}B_{j-1,j}}{B_{j-2,j}}\leq2\eta.
\end{equation*}
Set $\alpha_2=\frac{B_{j-2,j-1}B_{j-1,j}}{B_{j-2,j}}$, we can express $F_{j-2,j}$ as 
\begin{equation}
\label{equ:spj2}
F_{j-2,j}=B_{j-2,j}(1-\alpha_2)
\end{equation}
where $0<\alpha_2\ll1$ since the learning rate $\eta$ is usually a small number. Noting that $\alpha_2$ depends on the difference between $j$ and $j-2$ which is $2$ but not on $j$, we denote the fraction to be $\alpha_2$. From (\ref{equ:spj2}), we see that $F_{j-2,j}$ and $B_{j-2,j}$ would also be both large or small although there exists the factor $1-\alpha_2$.

For $i,j$ satisfying $i=j-3,j>3$, we get
\begin{equation}
\label{equ:spj3uf}
F_{j-3,j}=B_{j-3,j}[1-\alpha_{3,2}(1-\alpha_2)-\alpha_{3,1}]
\end{equation}
where $\alpha_{3,2}=\frac{B_{j-3,j-2}B_{j-2,j}}{B_{j-3,j}}$ and $\alpha_{3,1}=\frac{B_{j-3,j-1}B_{j-1,j}}{B_{j-3,j}}$, and thus $0<\alpha_{3,1},\alpha_{3,2}\ll1$. For $\alpha_2,\alpha_{3,1}$ and $\alpha_{3,2}$ all being small positive numbers, it is reasonable to have the following
\ifCLASSOPTIONonecolumn
\begin{equation}
\label{equ:approxj3}
1-\alpha_{3,2}(1-\alpha_2)-\alpha_{3,1}=1-\alpha_{3,1}-\alpha_{3,2}(1-\alpha_2)\approx(1-\alpha_{3,2})(1-\alpha_2).
\end{equation}
\else
\begin{align}
\label{equ:approxj3}
1-\alpha_{3,2}(1-\alpha_2)-\alpha_{3,1}=&1-\alpha_{3,1}-\alpha_{3,2}(1-\alpha_2)\nonumber\\
\approx&(1-\alpha_{3,2})(1-\alpha_2).
\end{align}
\fi
The approximation is because $\alpha_{3,1}$ and $\alpha_2$ are small positive numbers. It would be an equality instead when $\alpha_{3,1}=\alpha_2$. Similarly, $(1-\alpha_{3,2})(1-\alpha_2)$ can be considered as a square term because $\alpha_{3,2}$ and $\alpha_2$ are small positive numbers. In other words, we could choose $\alpha_3$ to satisfy
\begin{equation}
\label{equ:alpha3}
    1-\alpha_{3,2}(1-\alpha_2)-\alpha_{3,1}=(1-\alpha_3)^2
\end{equation}
and $\alpha_3$ is close to $\alpha_2,\alpha_{3,1}$ and $\alpha_{3,2}$. For example, in an experiment with $\eta=0.05$, it is possible to see $\alpha_2=0.08$, $\alpha_{3,1}=0.06$, and $\alpha_{3,2}=0.1$. Then, we should choose $\alpha_3=0.079$ which is around $\alpha_2,\alpha_{3,1}$ and $\alpha_{3,2}$ to satisfy (\ref{equ:alpha3}). Substituting (\ref{equ:alpha3}) back to (\ref{equ:spj3uf}), we get
\begin{equation}
F_{j-3,j}=B_{j-3,j}(1-\alpha_3)^2
\end{equation}
which implies that $B_{j-3,j}$ and $F_{j-3,j}$ are related with a factor $(1-\alpha_3)^2$.

Using mathematical induction we get the mathematical expression for the conformity property between $B_{i,j}$ and $F_{i,j}$ for $1\leq i<j\leq T+1$,
\begin{equation}
\label{equ:sp}
F_{i,j}=B_{i,j}(1-\alpha_{j-i})^{j-i-1}
\end{equation}
where $\alpha_{j-i}$ is a small positive number. 

The conformity property between $B_{i,j}$ and $F_{i,j}$ considers an exponential term, implying that when $j-i$ is small, it is easier to observe $F_{i,j}$ and $B_{i,j}$ to be large or small at the same time. However, when $j-i$ is large, one observes small $F_{i,j}$ values no matter what $B_{i,j}$ is. 

\section{Gaussian Variance for a Graph}
\label{sec:how}
\subsection{Impact of $\sigma^2$ on Predictions}
\label{subsec:sigma2ranges}
\begin{figure}[!t]
    \centering
    \includegraphics[width=8cm]{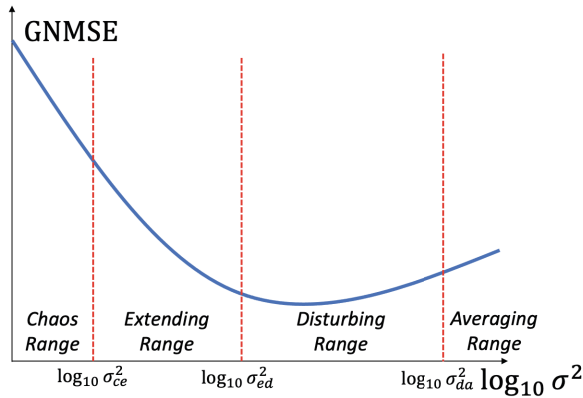}
    \caption{A typical GNMSE curve with respect to $\sigma^2$. The applicable interval of $\sigma^2$ is divided into four parts, i.e., the chaos range, the extending range, the disturbing range, and the averaging range, from left to right. The boundaries are denoted by $\sigma^2_{ce}$, $\sigma^2_{ed}$, and $\sigma^2_{ea}$, respectively.}
    \label{fig:ranges}
\end{figure}

We illustrate how prediction changes when $\sigma^2$ increases. Based on different behavior of $B_{i,T+1}$, $F_{i,T+1}$, we divide the possible range of $\sigma^2$, $(0,+\infty)$, into four parts, i.e., the Chaos Range, the Extending Range, the Disturbing Range, and the Averaging Range, as shown in Fig. \ref{fig:ranges}.

\subsubsection{Chaos Range}
In this range, $\sigma^2$ is so small that $B_{i,T+1}$ is close to $0$ when $\textbf{a}_i\neq\textbf{a}_{T+1}$. 

Following the conformity property, $F_{i,T+1}$ keeps pace with $B_{i,j}$. Note that in this case, $\alpha_{T+1-i}$ values are generally close to $0$. Take $\alpha_2\approx2\eta e^{-\frac{\|\textbf{d}_{j-2,j-1}\Vert^2+\|\textbf{d}_{j-1,j}\Vert^2-\|\textbf{d}_{j-2,j}\Vert^2}{2\sigma^2}}$ as an example. Because of the small $\sigma^2$ value, the exponent is a large negative number when the numerator of the exponent is nonzero. The small $\alpha_{T+1-i}$ values make the exponential term in (\ref{equ:sp}) decay slowly with decreasing $i$ from $T$, resulting in $F_{i,T+1}$ following $B_{i,T+1}$ closely. Like $B_{i,T+1}$, $F_{i,T+1}$ takes positive or negative values. Whereas positive $F_{i,T+1}$ values are viewed as weights of previously-seen nodal values contributing to the prediction, negative $F_{i,T+1}$ values play a disturbing role. Specifically, negative $F_{i,T+1}$ values cancel out positive $F_{i,T+1}$ with similar absolute values, resulting in taking nodal value difference instead of nodal values into consideration for predicting. Thus, we can find a minimum range which negative $F_{i,T+1}$ values fall in, and together with its positive counterpart, we get a symmetric range around $0$ which we call the \textit{noise range}. When $F_{i,T+1}$ takes value in the noise range, we say the corresponding nodes acquire an insignificant weight and do not contribute to the prediction.

Since most of $F_{i,T+1}$ values fall into the noise range when $\sigma^2$ is in the Chaos Range, the output is less predictable.

\subsubsection{Extending Range}
When $\sigma^2$ is in this range, $B_{i,T+1}$ values with small $\|\textbf{d}_{i,T+1}\Vert^2$ are significantly greater than $0$ whereas $B_{i,T+1}$ values with large $\|\textbf{d}_{i,T+1}\Vert^2$ are still close to $0$. 

\begin{figure}[!t]
    \centering
    \includegraphics[width=8cm]{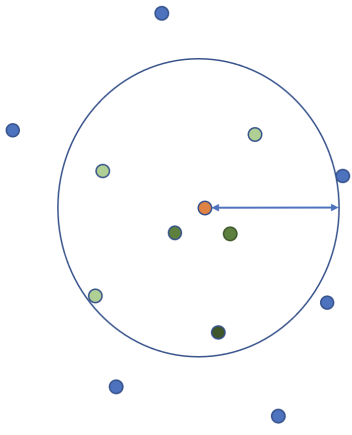}
    \caption{A simplified spacial illustration of sampled nodes and a tested node based on $\|\textbf{d}_{i,T+1}\Vert^2$. The orange node represents the tested node and the others are sampled nodes. The circle is centered at the tested node with the radius equal to the efficient distance. The different lightness of green nodes implies different weights.}
    \label{fig:distance}
\end{figure}

Considering the conformity between $B_{i,j}$ and $F_{i,j}$, it is expected that $F_{i,T+1}$ values for those small $\|\textbf{d}_{i,T+1}\Vert^2$ are significantly larger than $0$ while $F_{i,T+1}$ values for large $\|\textbf{d}_{i,T+1}\Vert^2$ are close to $0$. Due to the variation in $B_{i,T+1}$, the noise range still exists. However, there are $F_{i,T+1}$ values falling out of the noise range. We call $B_{i,T+1}$ whose corresponding $F_{i,T+1}$ falls outside the noise range the \textit{chosen $B_{i,T+1}$}. With the exponential approximation of $B_{i,j}$, we can find corresponding $\|\textbf{d}_{i,T+1}\Vert^2$ for chosen $B_{i,T+1}$ and have the maximum as the \textit{efficient distance}. Fig. \ref{fig:distance} shows a simplified spatial distribution for sampled nodes and a tested node. With the efficient distance as the radius, a circle centered at the tested node divides the sampled nodes into two groups. Then, we rewrite the prediction as
\begin{equation}
\label{equ:dividepreidction}
    \hat{f}_{T+1}
    =\sum_{\text{$F_{i,T+1}$ is significant}}y_iF_{i,T+1}+\sum_{\text{$F_{i,T+1}$ is insignificant}}y_iF_{i,T+1}.
\end{equation}
Sampled nodes inside the circle contribute to the first summation in (\ref{equ:dividepreidction}) with chosen $B_{i,T+1}$ and significant $F_{i,T+1}$. Sampled nodes outside the circle are assigned insignificant weights that fall into the noise range, so their nodal values contribute to the second summation which is less predictable. Clearly, if the sampled nodes inside the circle have nodal values close to that of the tested node, and if the second summation in (\ref{equ:dividepreidction}) is not dominant, the prediction would be close to its ground truth.

When $\sigma^2$ is increased within the Extending Range, the efficient distance grows. That is, the circle is extending to include more sampled nodes. This brings two benefits. First, more nodes are taken into consideration, instead of just a few nodes. Note that sampled nodes inside the circle have significant influence on prediction. When the circle includes only a few nodes all of which happen to have dissimilar nodal values with the tested node, the prediction would not be ideal. Enlarging the circle by increasing $\sigma^2$ within the Extending Range is helpful to include more sampled nodes, lowering weights of nodes with dissimilar nodal values. Second, fewer nodes take part into the less predictable part when the efficient distance grows. As a result, the performance of the SKG model gets better as $\sigma^2$ increases within the Extending Range. 

\subsubsection{Disturbing Range}
In this range, $B_{i,T+1}$ for all $\|\textbf{d}_{i,T+1}\Vert^2$ becomes significantly larger than $0$ but $B_{i,T+1}$ with small $\|\textbf{d}_{i,T+1}\Vert^2$ are significantly greater than $B_{i,T+1}$ with large $\|\textbf{d}_{i,T+1}\Vert^2$. Notice that it is the relative value not the absolute value of $B_{i,T+1}$ that carries information of similarity in adjacency vectors.

We can calculate an efficient distance using chosen $B_{i,T+1}$ values and draw a circle as in Fig. \ref{fig:distance}. However, the circle loses its role as a boundary. In fact, the circle includes most, if not all, of sampled nodes. Not all nodes inside the circle are assigned significant weights. Sampled nodes with higher $B_{i,T+1}$ still tend to get higher weights, but other sampled nodes would get significant weights if they show up at later times. Take the last sampled node in training as an example. It is assigned weight $F_{T,T+1}=B_{T,T+1}$ which is significantly greater than $0$. That is, the last nodal value is considered in prediction regardless of whether the node is spatially close to the tested node or not. In other words, predictions consider closeness in time in addition to similarity among adjacent vectors.

When $\sigma^2$ is increased within the Disturbing Range, $B_{i,T+1}$ values generally grow and predictions are focusing more and more on proximity of time. If the recent nodes do not happen to have similar nodal values with the tested node, the prediction will be far from its ground truth.

\subsubsection{Averaging Range}
In this range, $\sigma^2$ is so large that $B_{i,T+1}$ are close to $2\eta$ for all $\|\textbf{d}_{i,T+1}\Vert^2$.

Note that $\alpha_{T+1-i}$ values are close to $2\eta$ in this case. For example, $\alpha_2\approx2\eta e^{-\frac{\|\textbf{d}_{j-2,j-1}\Vert^2+\|\textbf{d}_{j-1,j}\Vert^2-\|\textbf{d}_{j-2,j}\Vert^2}{2\sigma^2}}\approx2\eta$. Consequently, $F_{i,T+1}$ is close to an exponential function with the base $1-2\eta$ as $i$ goes from $1$ to $T$.

Because of the exponential shape of $F_{i,T+1}$ with respect to $i$, it is unsurprising that only recent nodes are taken into consideration in predicting. Besides, it is the same set of sampled nodes that take significant weights in calculation for different tested nodes, and $\{F_{i,T+1}\}_{i=1}^T$ are similar for different tested nodes. Thus, it is anticipated that outputs of the model are about the same for all tested nodes.

\subsection{How to Choose a $\sigma^2$}
For clarity, let us denote the boundary between the Chaos Range and the Extending Range by $\sigma_{ce}^2$, the boundary between the Extending Range and the Disturbing Range by $\sigma_{ed}^2$, and the boundary between the Disturbing Range and the Averaging Range by $\sigma_{da}^2$. From the analysis in Section \ref{subsec:sigma2ranges}, we conclude that the performance is bad in the Chaos Range, gets better in the Extending Range, might get worse in the Disturbing Range, and is bad in the Averaging Range. As predictions in the Disturbing Range consider more proximity of time instead of network topology than in the Extending Range, we choose the boundary between the Extending Range and the Disturbing Range $\sigma_{ed}^2$ as a suitable $\sigma^2$ (cases where performance gets the best in the Disturbing Range are discussed in Section \ref{subsubsec:algorithmperformance}). Using Fig. \ref{fig:distance} as an illustration, the radius of the circle achieves its maximum while not including nodes with dissimilar adjacency vectors on this boundary. Intuitively speaking, what happens at $\sigma_{ed}^2$ is $B_{i,T+1}$ values for large $\|\textbf{d}_{i,T+1}\Vert^2$ are ``just significantly greater than $0$." 

We should find the largest possible $\|\textbf{d}_{i,T+1}\Vert^2$ value. However, we cannot know the distribution of $\|\textbf{d}_{i,T+1}\Vert^2$ when we configure the SKG model. So, we use the largest value of $\|\textbf{d}_{i,j}\Vert^2$ among all pairs of sampled nodes, denoted by $\|\textbf{d}\Vert^2_{max}$, instead. We also need to give a concrete math expression for ``significantly greater than $0$." Recall that $F_{i,T+1}$ is considered as significant if it falls out of the noise range. Following the conformity property between $B_{i,j}$ and $F_{i,j}$, $F_{i,T+1}$ is likely to fall out of the noise range when $B_{i,T+1}$ falls out of the noise range. Then, $B_{i,T+1}$ is significant when it is greater than the upper bound of the noise range. 

It can be observed that the noise range exists for different $\sigma^2$ values. The existence is (partly) due to the variation of $B_{i,T+1}$ for $1\leq i\leq T$. For example, in an extreme case where $B_{i,T+1}=2\eta$ for $1\leq i\leq T$, $F_{i,T+1}$ is an exact exponential function with respect to $i$ and the noise range vanishes. Detailed analysis on the noise range is left for future study. Although the noise range changes with $B_{i,T+1}$ variation, the change is limited. So, we can calculate the noise range in the Chaos Range which is easier to derive and use it for all $\sigma^2$. Recall in the Chaos Range, $F_{i,T+1}$ values closely follow corresponding $B_{i,T+1}$ values. From (\ref{equ:var}), it is known that 
\begin{equation*}
    \Var[B_{i,j}|\textbf{d}_{i,j}]=\frac{(2\eta)^2}{2D}\Big(e^{-\frac{\|\textbf{d}_{i,j}\Vert^2}{\sigma^2}}-1\Big)^2
\end{equation*}
which is a decreasing function with respect to $\|\textbf{d}_{i,j}\Vert^2$. That is, $\Var[B_{i,T+1}]$ which is an upper bound of $\Var[F_{i,T+1}]$ is upper-bounded by $\Var[B_{i,T+1}|\textbf{0}]\leq\frac{(2\eta)^2}{2D}$. Then, the upper bound of the noise range $noise_{up}$ can be approximated by the standard deviation as
\begin{equation}
    \label{equ:noiseup}
    noise_{up}\approx\sqrt{\frac{(2\eta)^2}{2D}}=\frac{1}{\sqrt{2D}}\times2\eta.
\end{equation}
For a more precise noise range at the boundary between the Extending Range and the Disturbing Range, please follow \textbf{Algorithm \ref{alg:findnoiserange}}. 

Once making sure $noise_{up}$ and $\|\textbf{d}\Vert^2_{max}$, we can apply the exponential approximation of $B_{i,T+1}$ and get
\begin{equation*}
    2\eta e^{-\frac{\|\textbf{d}\Vert^2_{max}}{2\sigma_{ed}^2}}=noise_{up}
\end{equation*}
which is equivalent to
\begin{equation}
\label{equ:edthresh}
    \sigma_{ed}^2=-\frac{\|\textbf{d}\Vert^2_{max}}{2\ln{\frac{noise_{up}}{2\eta}}}.
\end{equation}
The steps of how to choose a suitable $\sigma^2$ are summarized in \textbf{Algorithm \ref{alg:choosesigma2}}. We would like to mention that, although (\ref{equ:noiseup}) and (\ref{equ:edthresh}) indicate that $D$ affects calculated $\sigma_{ed}^2$, the best $\sigma^2$ is not influenced by $D$ theoretically as long as $D$ is sufficiently large. Note that $D$ cannot be arbitrarily small for the validity of the random feature approximation. The proposed $\sigma^2$ is close to the optimal one, but not exactly the same. In this case, (\ref{equ:noiseup}) provides an approximation of $noise_{up}$ and we also provide \textbf{Algorithm
\ref{alg:findnoiserange}} to mitigate the impact of $D$ on the proposed $\sigma^2$.

\begin{algorithm}[H]
\caption{Choosing $\sigma^2$ for the Gaussian Kernel in SKG}\label{alg:choosesigma2}
\begin{algorithmic}
\STATE
\STATE \textbf{Input:} adjacency vectors for all sampled nodes, and the number of the random features $D$.
\STATE Get $noise_{up}$ via (\ref{equ:noiseup}) (alternatively, for a more precise noise range, use \textbf{Algorithm \ref{alg:findnoiserange}});
\STATE Get $\|\textbf{d}_{i,j}\Vert^2$ for all pairs of sampled nodes and record the maximum value;
\STATE Get a $\sigma^2$ value via (\ref{equ:edthresh});
\end{algorithmic}
\label{choosesigma2}
\end{algorithm}

\begin{algorithm}[H]
\caption{Finding a More Precise Noise Range}\label{alg:findnoiserange}
\begin{algorithmic}
\STATE Run a simulation with $\sigma^2$ found with $noise_{up}$ in (\ref{equ:noiseup});
\STATE Get $F_{i,T+1}$ via (\ref{equ:fijdef}) for $1\leq i\leq T$ and record its minimum;
\STATE Using the absolute value of the minimum as the new $noise_{up}$, the new noise range is $[-noise_{up},noise_{up}]$;
\end{algorithmic}
\label{findnoiserange}
\end{algorithm}

We can have similar definitions for the boundary between the Chaos Range and the Extending Range  $\sigma_{ce}^2$ and the boundary between the Disturbing Range and the Averaging Range $\sigma_{da}^2$, which are used in later simulations. The boundary $\sigma_{ce}^2$ should be such that $B_{i,T+1}$ for the smallest nonzero $\|\textbf{d}_{i,T+1}\Vert^2$ is greater than $noise_{up}$. With the exponential property, we get
\begin{equation*}
    2\eta e^{-\frac{\|\textbf{d}\Vert^2_{min,nonzero}}{2\sigma_{ce}^2}}=noise_{up}
\end{equation*}
where $\|\textbf{d}\Vert^2_{min,nonzero}$ denotes the smallest nonzero $\|\textbf{d}_{i,T+1}\Vert^2$, or equivalently,
\begin{equation}
\label{equ:cethresh}
    \sigma_{ce}^2=-\frac{\|\textbf{d}\Vert^2_{min,nonzero}}{2\ln{\frac{noise_{up}}{2\eta}}}.
\end{equation}
The boundary $\sigma_{da}^2$ should result in $B_{i,T+1}$ for the largest $\|\textbf{d}_{i,T+1}\Vert^2$ to be close to $2\eta$. That is, $\sigma_{da}^2$ satisfies
\begin{equation*}
    2\eta e^{-\frac{\|\textbf{d}\Vert^2_{max}}{2\sigma_{da}^2}}=2\eta(1-closeness)
\end{equation*}
or equivalently,
\begin{equation}
\label{equ:dathresh}
    \sigma_{da}^2=-\frac{\|\textbf{d}\Vert^2_{max}}{2\ln(1-closeness)}.
\end{equation}
where $closeness$ should be chosen as a small value which indicates how much the value is expected to be close to $2\eta$. For example, $closeness$ has to be within $(0,0.5)$ to imply the value is closer to $2\eta$ than $0$. The choice is somewhat arbitrary as long as it indicates nearness to $2\eta$.

\section{Simulations}
\label{sec:sim}
In this section, we provide simulation results confirming some properties of $B_{i,j}$ and $F_{i,j}$, and show the performance of the proposed algorithm on four real datasets.

\subsection{Performance Measure for SKG}
When talking about the performance of SKG, we follow \cite{OnlineGradraker} and use \textit{generalization normalized mean squared error (GNMSE)} as the metric. GNMSE is defined as
\begin{equation}
\label{equ:gnmse}
\text{GNMSE}=\frac{\|\textbf{y}_{true}-\textbf{y}_{pred}\Vert^2}{\|\textbf{y}_{true}\Vert^2}
\end{equation}
where $\textbf{y}_{pred}$ and $\textbf{y}_{true}$ are vectors whose elements are the predicted and the true nodal values for all tested nodes, respectively. Besides, we will use normalized true values and predictions to calculate GNMSE.

\subsection{Real Datasets}
\label{subsec:casestudy}
We use four real datasets, the Temperature-Jan dataset, the Cora-Con dataset, and the Email-EU-Core dataset.

\subsubsection{The Temperature-Jan Dataset}
The Temperature-Jan dataset is a part of the Temperature dataset. The Temperature dataset contains the average monthly temperature information of 93 weather stations during 1961-1990 \cite{TemperatureDataset1} and that of 91 weather stations during 1981-2010 \cite{TemperatureDataset2} in Switzerland. We take the intersection of the two sets of stations and get $83$ stations. We view the 83 weather stations to be nodes. Note that the Temperature dataset does not contain any graph. It has altitude information of the stations. 
We used the altitude information to create two graphs. The first graph was created in the same way as authors of \cite{LearnLaplacian} created their ground truth graph based on the altitudes of stations. That is, an edge exists between a pair of nodes only when the altitude difference between the corresponding stations is less than $300$ meters. The second graph is the same except that weights of connected nodes are not $1$ but follow $e^{-\Delta/300}$ where $\Delta$ corresponds to the absolute value of the altitude difference between a pair of connected nodes. The Temperature-Jan dataset contains the monthly average temperature information of all the $83$ stations in January during 1961-1990, and the created graph of the $83$ stations during the same period. Note that the first graph is unweighted whereas the second one is weighted.
The temperature for the stations is viewed as nodal values. 

\subsubsection{The Cora-Con Dataset}
The Cora-Con dataset is part of the Cora dataset. The Cora dataset \cite{CoraDataset} contains a citation network of $2708$ scientific papers each of which is categorized as one of seven topics in the field of machine learning. We view the papers as nodes. Note that the citation network is an unweighted directed graph where edges can point from a citing paper to a cited paper. Then, the (column) adjacency vector of a node is actually an indicator vector of whether the paper cites a list of papers. We assign an integer from $\{1,2,...,7\}$ representing paper classes as nodal values. The Cora dataset contains $486$ papers with no citing. That is, these nodes have the adjacency vector of $\textbf{0}$. But they carry different nodal values. To avoid these nodes confusing the SKG model, we create the Cora-Con dataset by excluding the $486$ nodes, remaining $2222$ nodes and the related network.

\subsubsection{The Email-EU-Core Dataset}
The Email-EU-Core dataset \cite{eec} contains email communication among 1005 members in a European research institution. Every member belongs to one of $42$ departments. We view the members as nodes and assign an integer from $\{1,2,...,42\}$ representing their membership as their nodal values. The communication network is unweighted and directed.

\subsubsection{The Wikipedia-Math-Daily Dataset}
The Wikipedia-Math-Daily dataset is part of the Wikipedia-Math dataset \cite{wikiMath}. The Wikipedia-Math dataset contains a weighted link network among $1068$ Wikipedia pages about Mathematics topics. The web pages are viewed as nodes and the network is directed. Weights on the links denote relevance. The dataset also contains daily visits for those pages between $2019$ and $2021$ March, $731$ days in total. The daily visits on any day can be used as ground truth. The Wikipedia-Math-Daily contains the directed weighted network and daily visits on March 16th, 2019.

\subsection{Exponential Approximation of $B_{i,j}$}
The exponential approximation property is one of the core assumptions for other properties of $B_{i,j}$ and $F_{i,j}$. We aim to compare the exponential approximation with practical distributions of $B_{i,j}$. The Temperature-Jan dataset is used where $40\%$ of total nodes ($33$ nodes) are randomly selected as sampled nodes. Referencing nodes are the sampled nodes. The parameter $D$ is $200$. The learning rate $\eta$ is set to $0.1$. The Gaussian variance $\sigma^2$ is set to $10$.

Following the SKG algorithm and the definition of $B_{i,j}$ in (\ref{equ:bijdef}), $B_{i,j}$ values for all pairs of sampled nodes are calculated. We select $B_{i,j}$ values with $\|\textbf{d}_{i,j}\Vert^2=15$ as an example and get their distribution. There are $21$ qualified pairs of nodes and the corresponding similarity measure take values within $(0.085,0.105)$. The sample mean is $0.0976$, and the sample variance is $5.80\times 10^{-5}$. Recall that the exponential approximation in (\ref{equ:eb}) states that $B_{i,j}$ can be approximated as $2\eta e^{-\frac{\|\textbf{d}_{i,j}\Vert^2}{2\sigma^2}}=0.094$ when $\|\textbf{d}_{i,j}\Vert^2=15$. Comparing with the potential range $(0,0.2]$ for $B_{i,j}$ with no $\|\textbf{d}_{i,j}\Vert^2$ constraints, we could say the exponential approximation is valid.

\subsection{Conformity Property and the Impact of $\sigma^2$}
The conformity property is at the core of the analysis of the impact of $\sigma^2$. We show $B_{i,j}$ and $F_{i,j}$ behavior with $\sigma^2$ in different ranges using the Temperature-Jan dataset. Again, $40\%$ of total nodes are randomly chosen as sampled nodes which are also referencing nodes. The parameters are $D=200$, $\eta=0.1$, and $E=3$.

We first make sure the boundaries between adjacent ranges. Checking with the Temperature-Jan dataset, we know $\|\textbf{d}\Vert^2_{min,nonzero}=1$, and $\|\textbf{d}\Vert^2_{max}=27$. With $noise_{up}$ in (\ref{equ:noiseup}) and $closeness=0.1$, we get $\sigma_{ce}^2=0.22$, $\sigma_{ed}^2=5.87$, and $\sigma_{da}^2=135$. Fig. \ref{fig:sigma2impact} shows examples of $B_{i,T+1}$ and $F_{i,T+1}$ with $\sigma^2$ in different ranges. The conformity between $B_{i,j}$ and $F_{i,j}$ can be observed from the figures. Additionally, Fig. \ref{fig:alpha10} displays an example of $\alpha_{T+1-i}$ values when $\sigma^2$ is in its Disturbing Range.

\ifCLASSOPTIONonecolumn
\begin{figure}[!t]
\centering
\subfloat[]{\includegraphics[width=8cm]{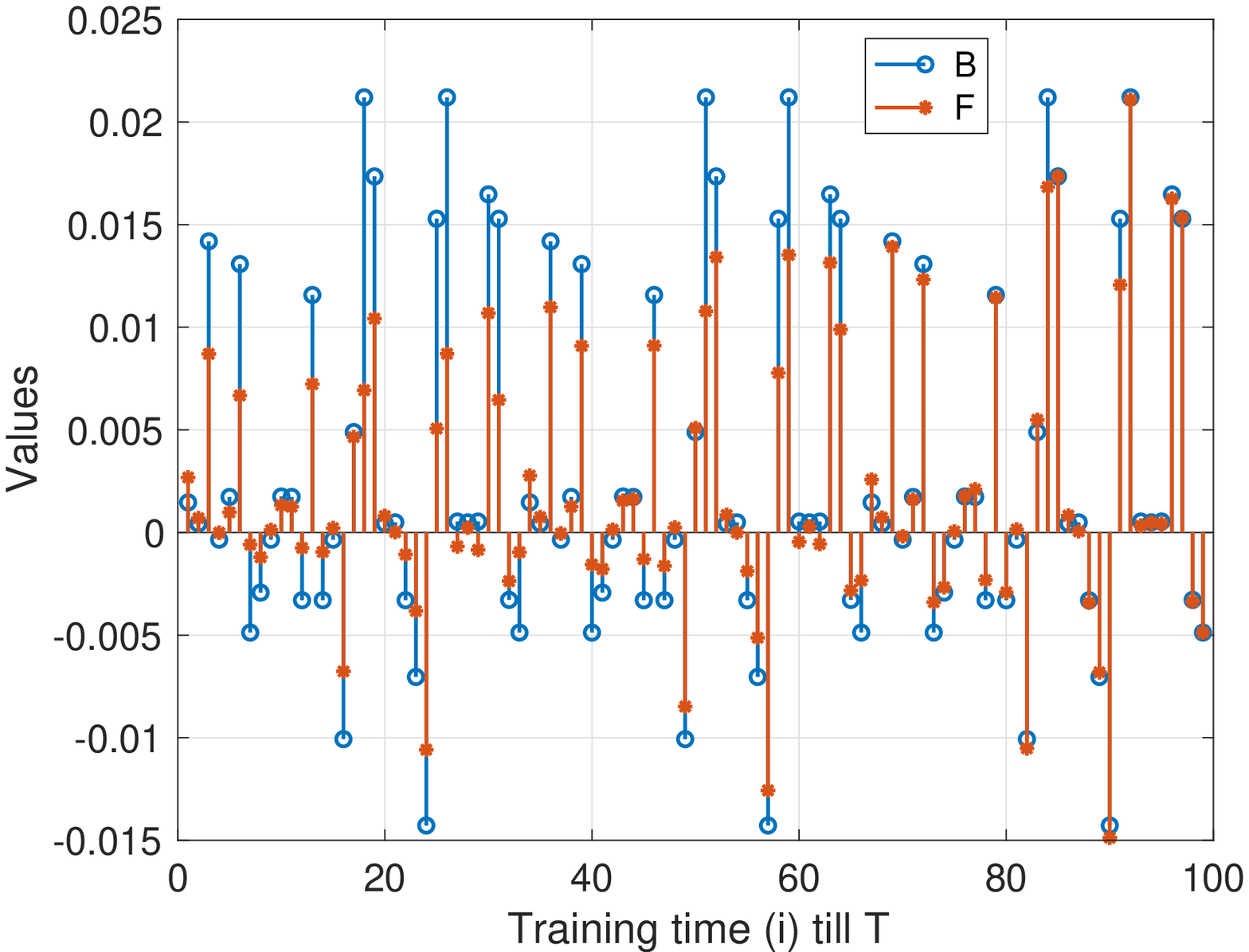}}
\subfloat[]{\includegraphics[width=8cm]{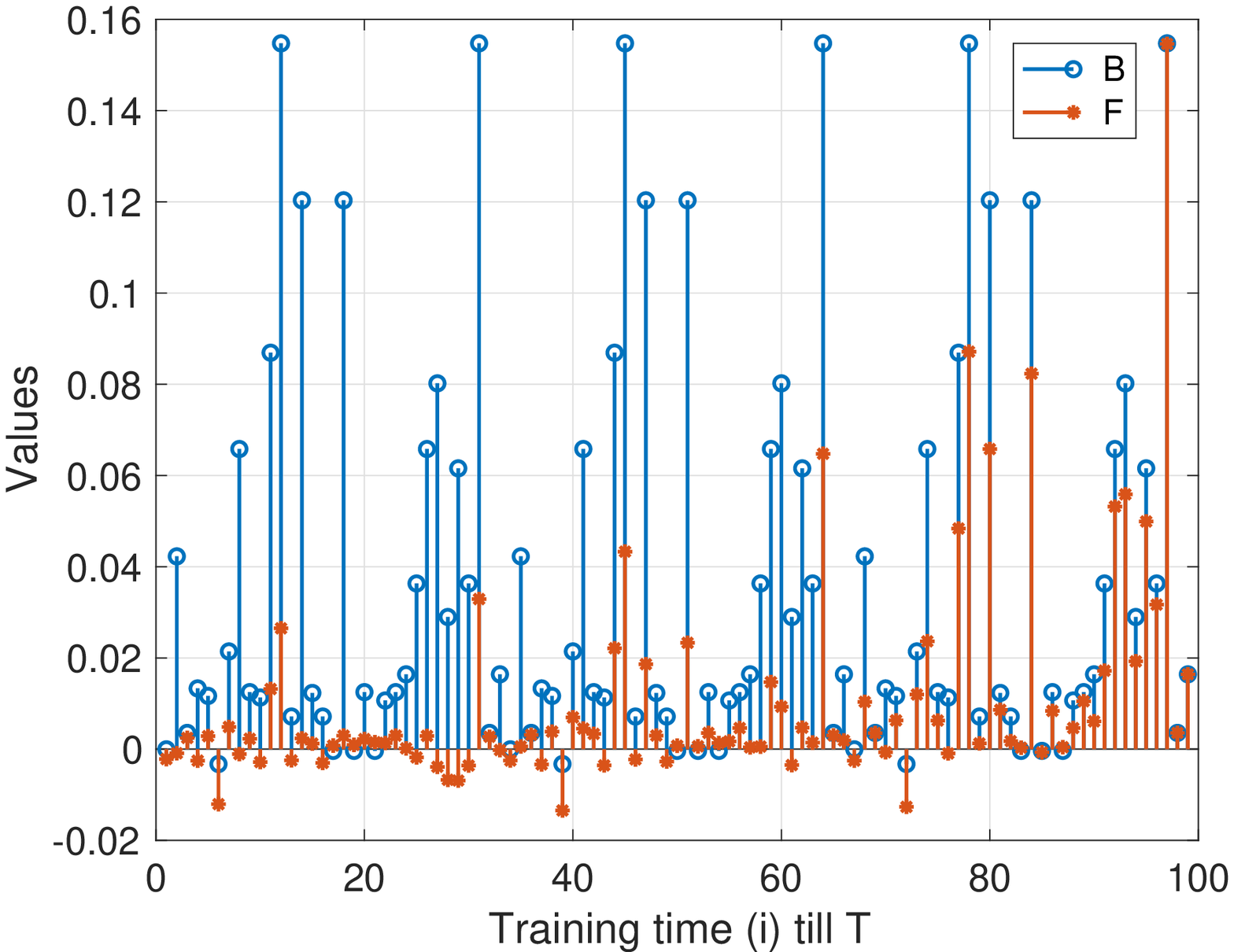}}
\hfill
\subfloat[]{\includegraphics[width=8cm]{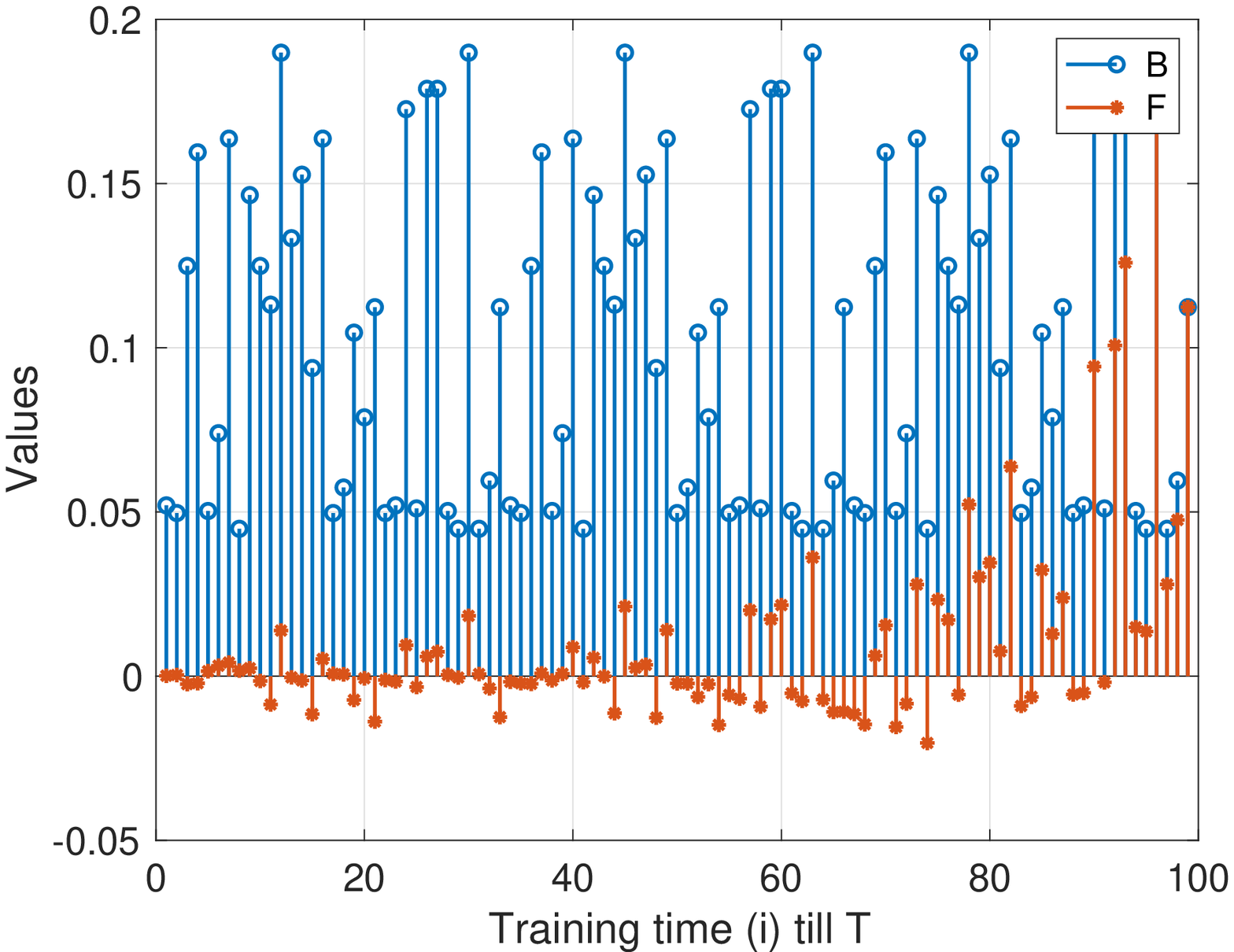}}
\subfloat[]{\includegraphics[width=8cm]{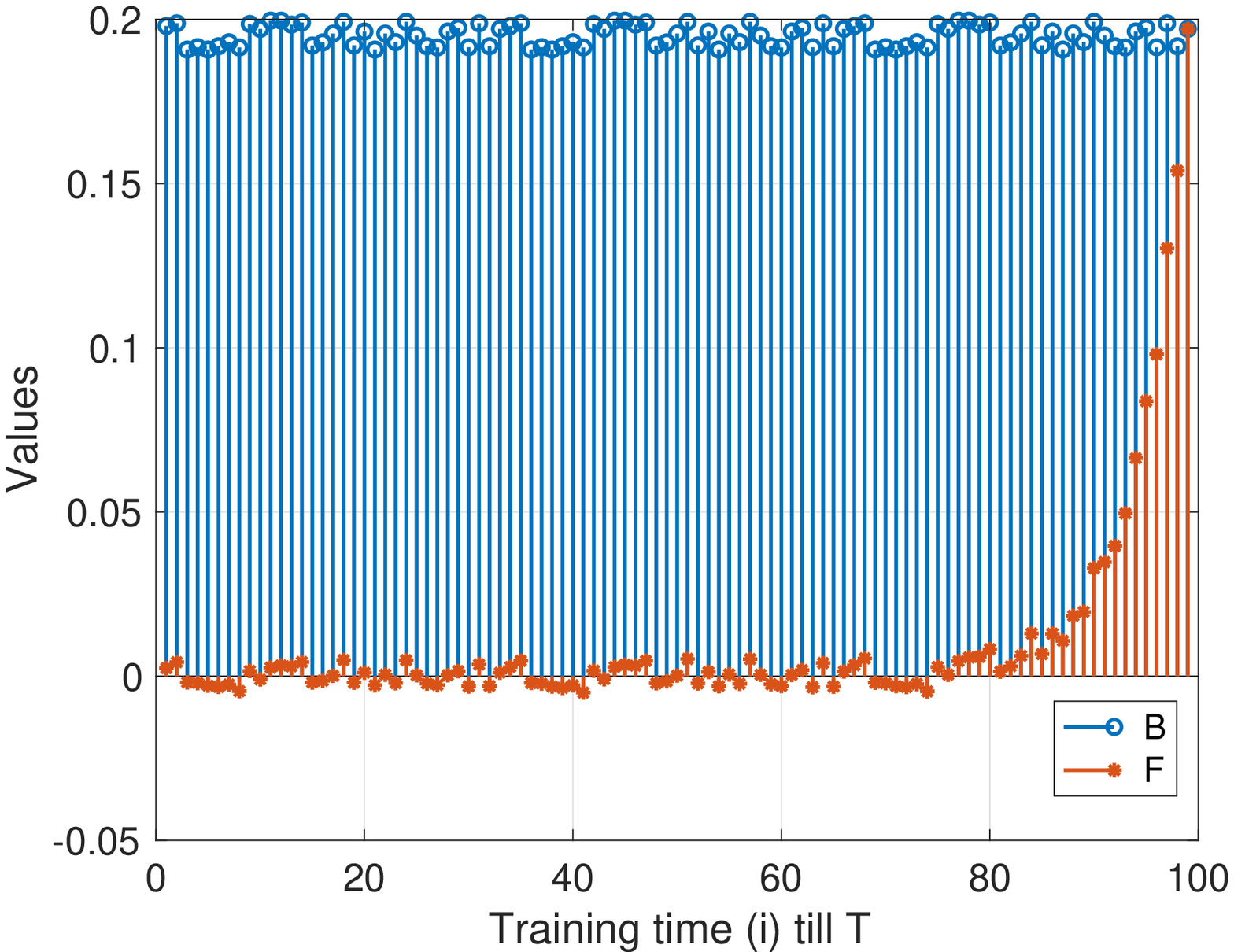}}
\caption{Values of $B_{i,T+1}$ and $F_{i,T+1}$ for a tested node in the Temperature-Jan dataset with different $\sigma^2$ values. (a) $\sigma^2=0.1$. (b) $\sigma^2=2$. (c) $\sigma^2=10$. (d) $\sigma^2=300$.}
\label{fig:sigma2impact}
\end{figure}

\begin{figure}[!t]
    \centering
    \includegraphics[width=8cm]{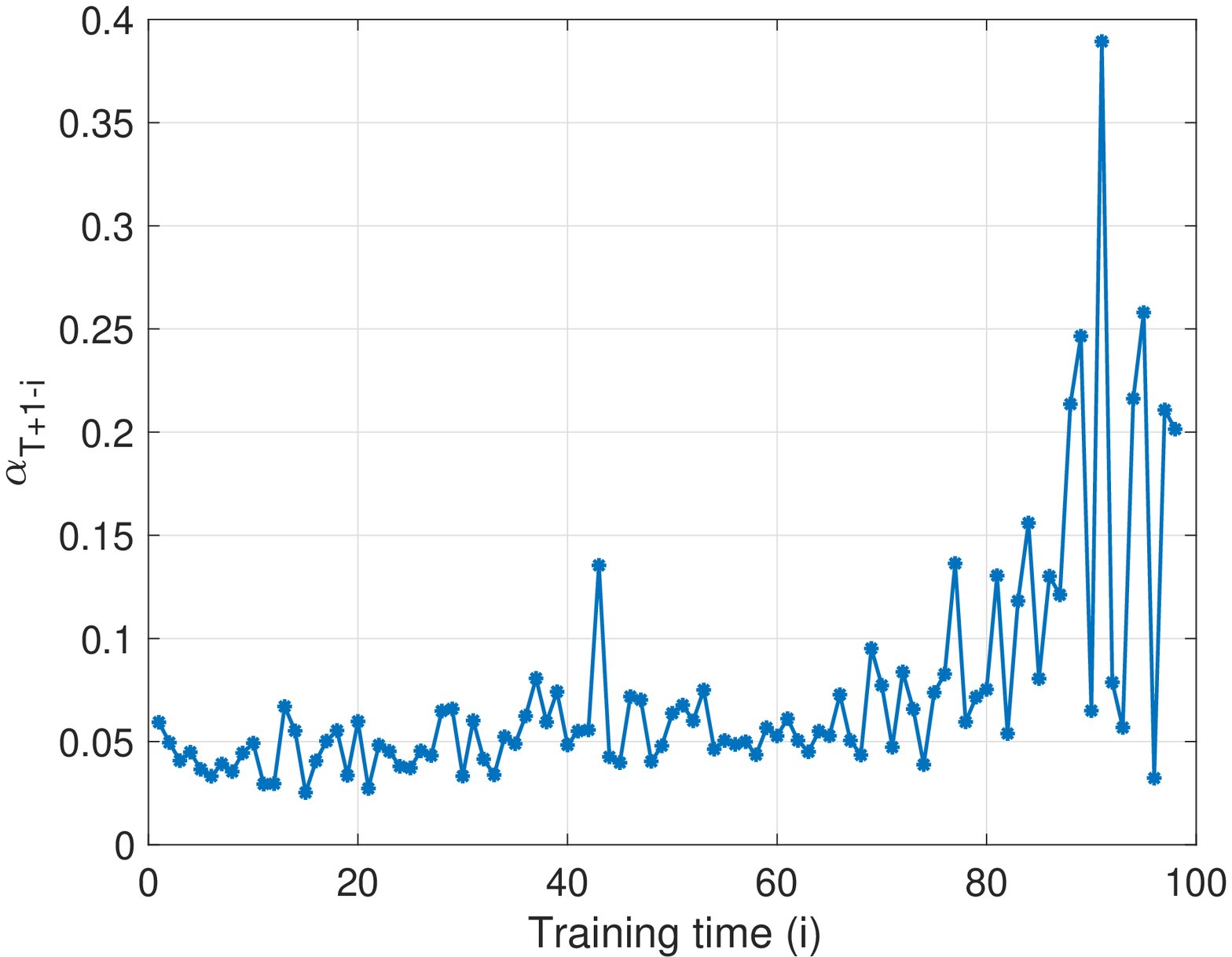}
    \caption{Values of $\alpha_{T+1-i}$ for a tested node in the Temperature-Jan dataset with $\sigma^2=10$.}
    \label{fig:alpha10}
\end{figure}
\begin{figure}[!t]
    \centering
    \includegraphics[width=8cm]{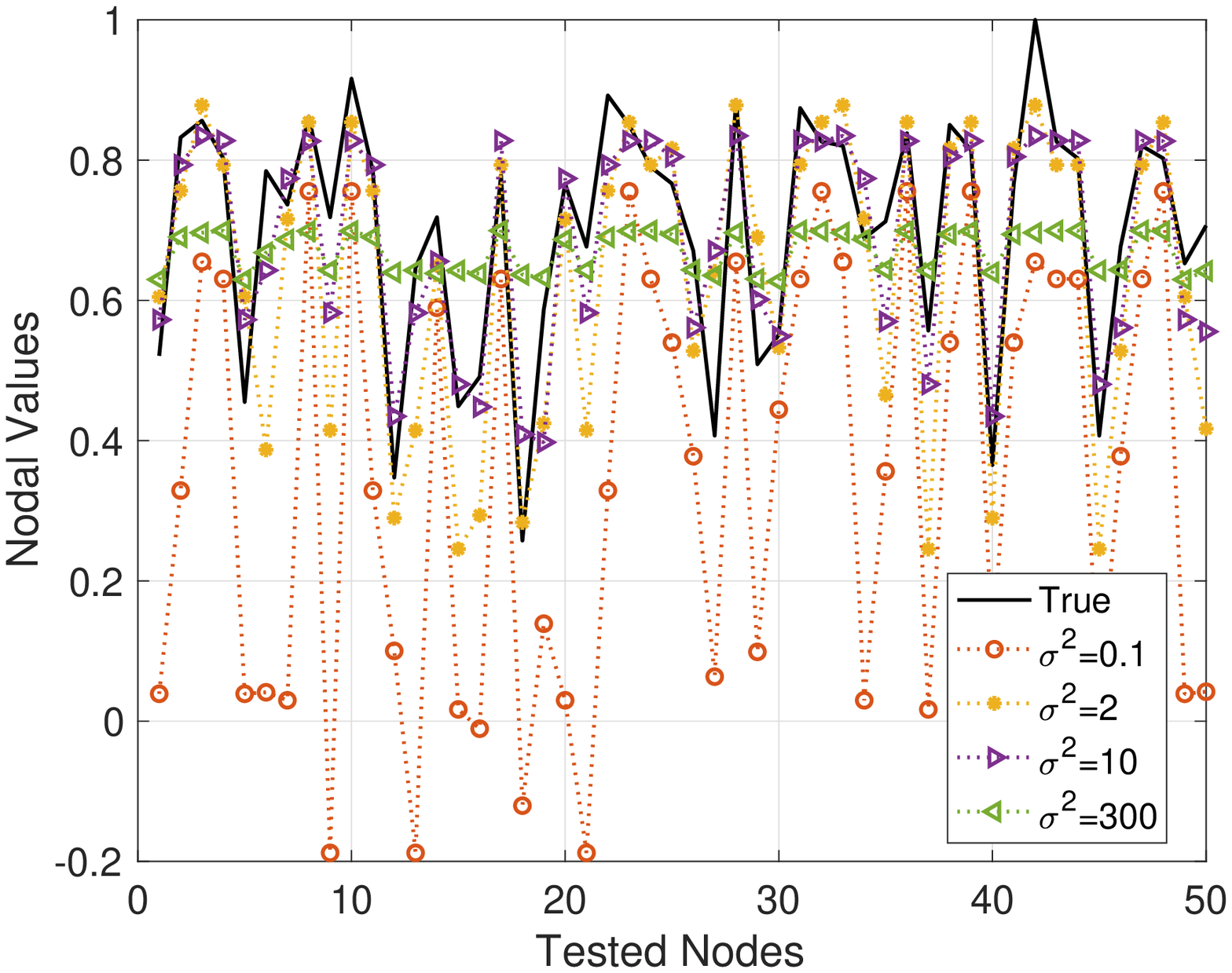}
    \caption{True nodal values and predicted values when $\sigma^2=0.1$, $\sigma^2=2$, $\sigma^2=10$, and $\sigma^2=300$ for tested nodes in the Temperature-Jan dataset.}
    \label{fig:temperaturejanpred}
\end{figure}

\else

\begin{figure}[!t]
\centering
\subfloat[]{\includegraphics[width=4.5cm]{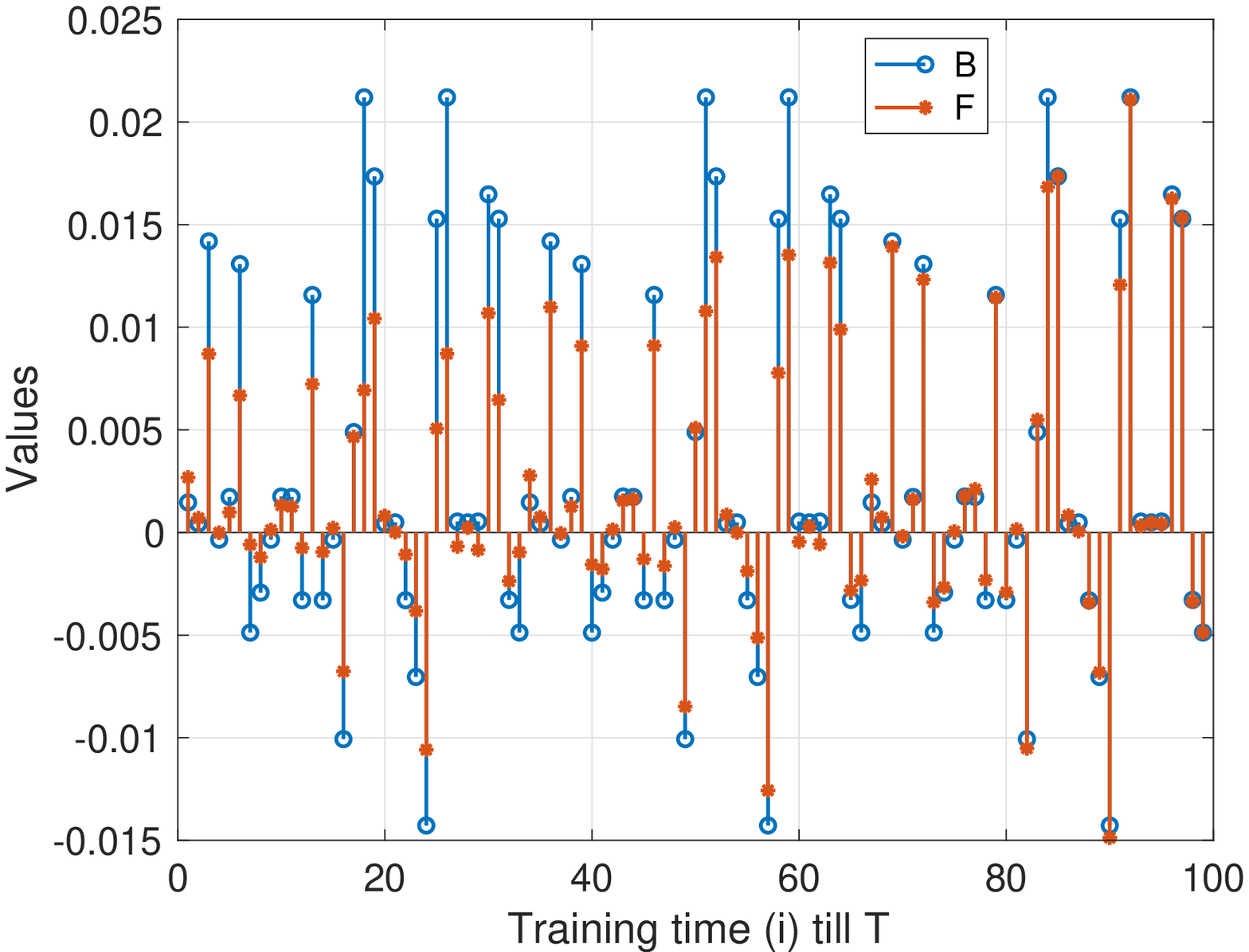}}
\subfloat[]{\includegraphics[width=4.5cm]{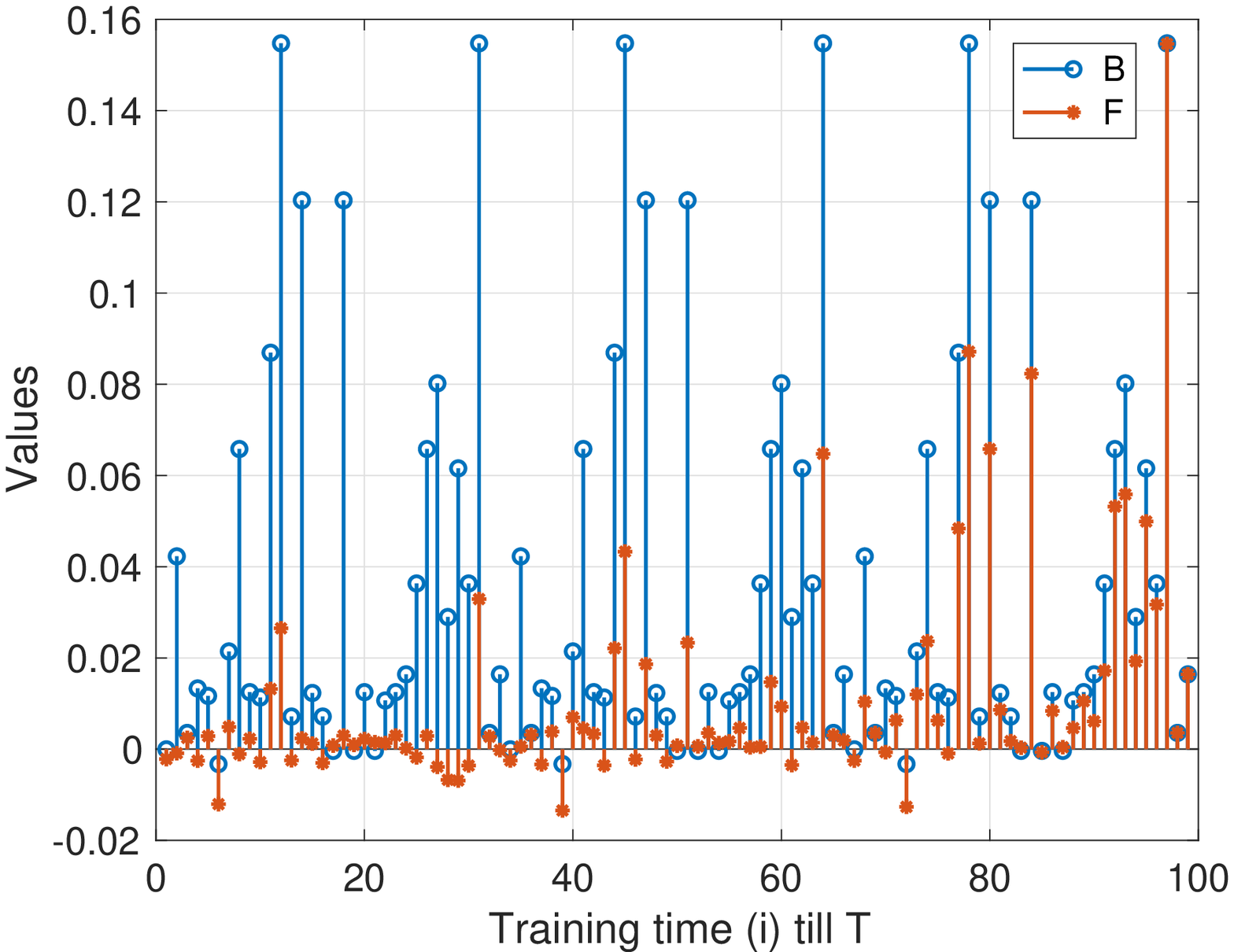}}
\hfill
\subfloat[]{\includegraphics[width=4.5cm]{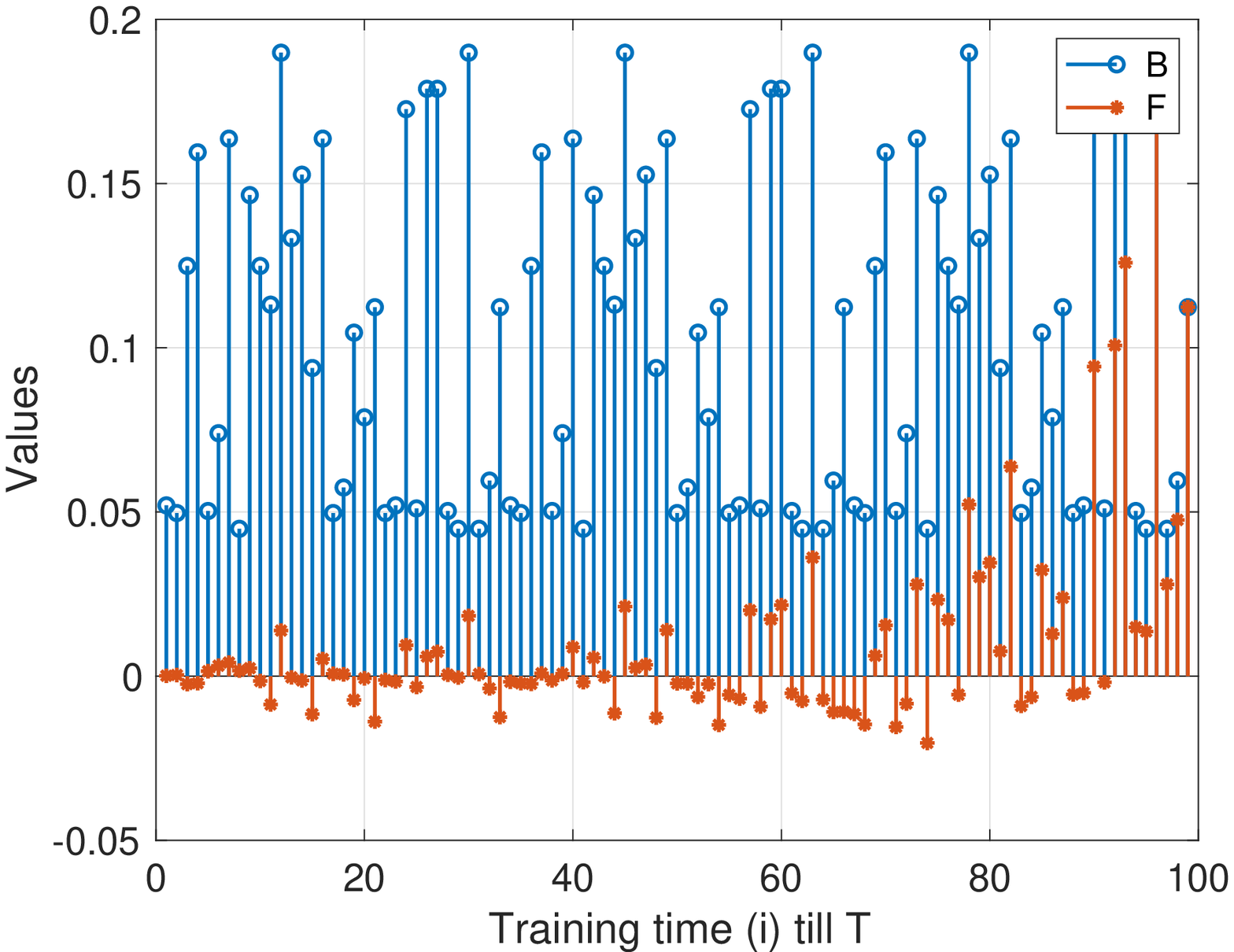}}
\subfloat[]{\includegraphics[width=4.5cm]{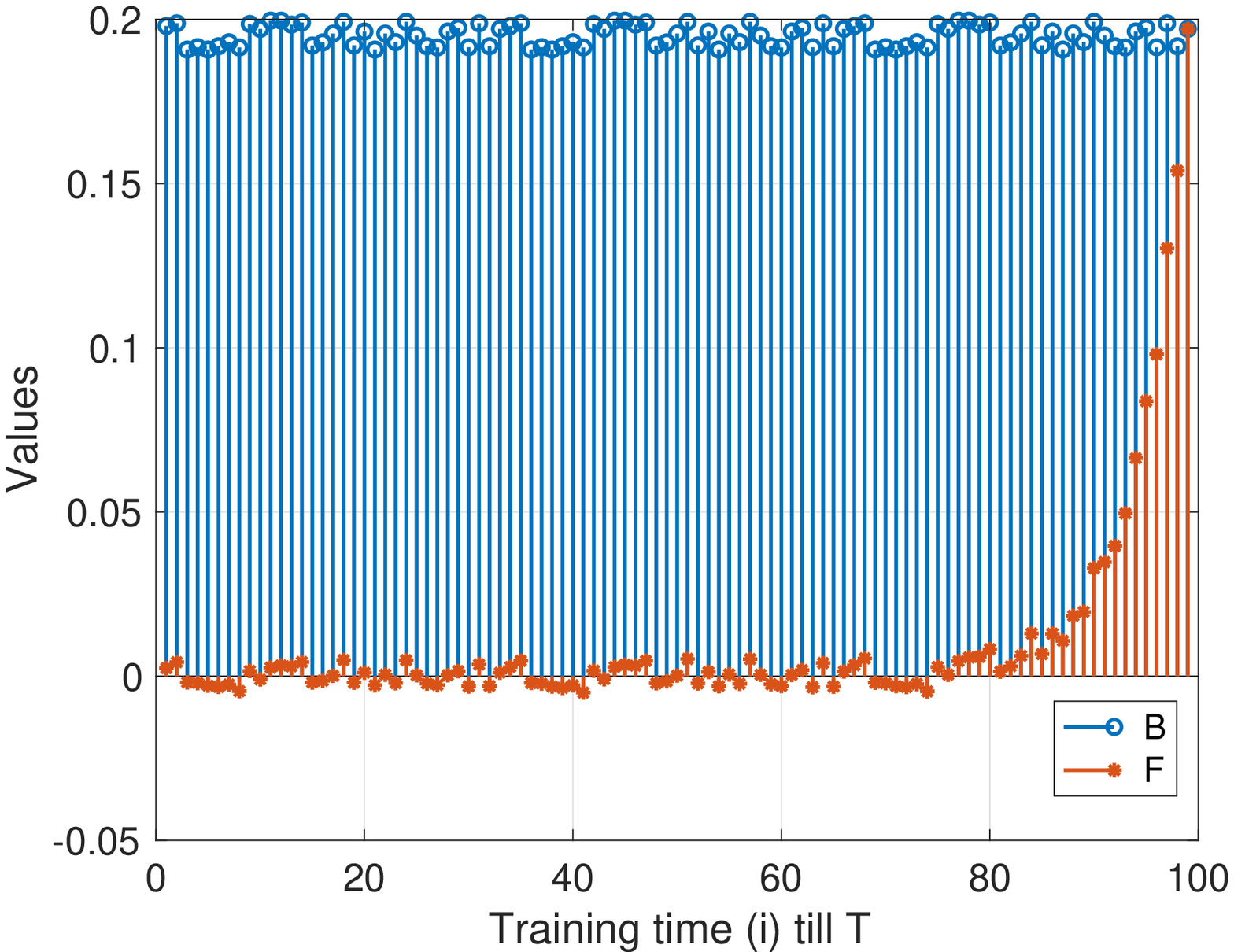}}
\caption{Values of $B_{i,T+1}$ and $F_{i,T+1}$ for a tested node in the Temperature-Jan dataset with different $\sigma^2$ values. (a) $\sigma^2=0.1$. (b) $\sigma^2=2$. (c) $\sigma^2=10$. (d) $\sigma^2=300$.}
\label{fig:sigma2impact}
\end{figure}

\begin{figure}[!t]
    \centering
    \includegraphics[width=8cm]{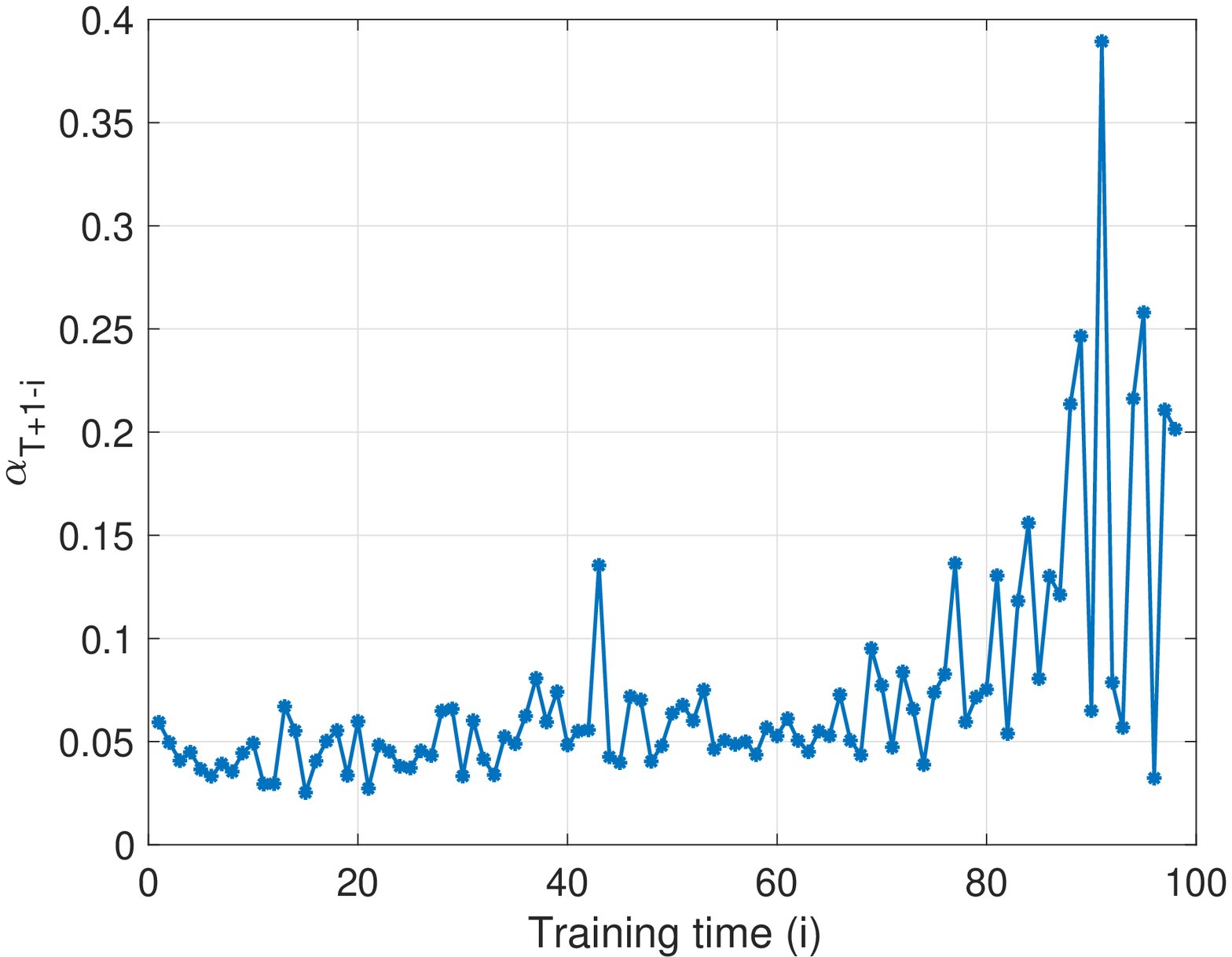}
    \caption{Values of $\alpha_{T+1-i}$ for a tested node in the Temperature-Jan dataset with $\sigma^2=10$.}
    \label{fig:alpha10}
\end{figure}

\begin{figure}[!t]
    \centering
    \includegraphics[width=8cm]{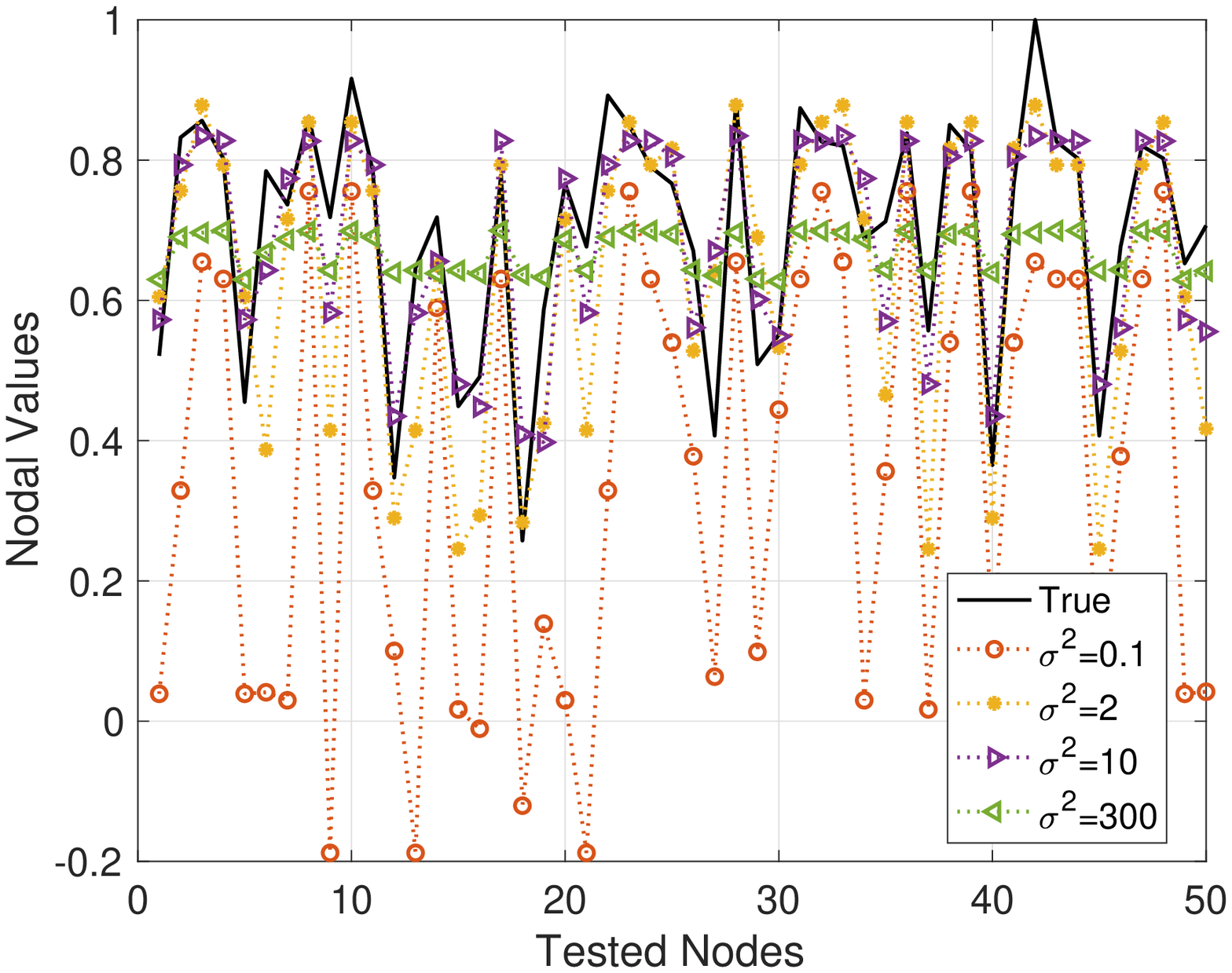}
    \caption{True nodal values and predicted values when $\sigma^2=0.1$, $\sigma^2=2$, $\sigma^2=10$, and $\sigma^2=300$ for tested nodes in the Temperature-Jan dataset.}
    \label{fig:temperaturejanpred}
\end{figure}
\fi

Given detailed look, it is seen from Fig. \ref{fig:sigma2impact}(a) that when $\sigma^2$ is in its Chaos Range, $B_{i,T+1}$ values are around $0$, and so do $F_{i,T+1}$ values. Fig. \ref{fig:sigma2impact}(b) verifies that when $\sigma^2$ is in its Extending Range, positive and negative values of $B_{i,T+1}$ become unbalanced, and some $F_{i,T+1}$ values are greatly larger than $0$. The $B_{i,T+1}$ values are greater than $0$ in Fig. \ref{fig:sigma2impact}(c), and $F_{i,T+1}$ for the penultimate node are relatively large although its $B_{i,T+1}$ is close to $\min\{B_{i,T+1}\}_{i=1}^T$. In Fig. \ref{fig:sigma2impact}(d), $B_{i,T+1}$ values are all close to $2\eta$, and $F_{i,T+1}$ is roughly an exponential function with respect to $i$.

Fig. \ref{fig:temperaturejanpred} plots predictions for the tested nodes under different $\sigma^2$ as well as their true nodal values. Predictions with $\sigma^2=0.1$ have the greatest error whereas predictions with $\sigma^2=300$ are almost the same. In summary, as $\sigma^2$ grows, predictions tend to get closer to their ground truth, while $\sigma^2$ grows furthermore than needed, predictions are roughly the same for different tested nodes for the Temperature-Jan dataset.

\subsection{Performance of the Proposed Algorithm}
To show the performance of the proposed algorithm, we compare the theoretical $\sigma_{ed}^2$ value from \textbf{Algorithm \ref{alg:choosesigma2}} with the best $\sigma^2$ found by simulations using the three real datasets. In each dataset, $40\%$ of total nodes are randomly selected as the sampled nodes which are also referencing nodes. The $noise_{up}$ in (\ref{equ:noiseup}) of $F_{i,j}$ is used. Simulation results will be denoted in blue solid curves and their corresponding proposed $\sigma^2_{ed}$ will be denoted in red dotted line.

For unweighted graph in the Temperature-Jan dataset, the value of $\|\textbf{d}\Vert^2_{max}$ is found to be $27$, and the theoretical result is $\sigma_{ed}^2=5.86$. For the weighted graph, the value of $\|\textbf{d}\Vert^2_{max}$ is found to be $12.49$ resulting in the theoretical result $\sigma_{ed}^2=2.08$. When finding the relationships between GNMSE and $\sigma^2$ by simulations, we set $E=3$, $D=200$, and $\eta=0.1$. The results are shown in Fig. \ref{fig:temperaturejan} and Fig. \ref{fig:temperatureweighted}, respectively. Note that the shown values of GNMSE are averaged over $50$ repeated experiments. 

For the Cora-Con dataset, the value of $\|\textbf{d}\Vert^2_{max}$ is found out to be $8$, resulting a theoretical value $\sigma_{ed}^2=1.05$. For simulations, we set $E=3$, $D=M=888$, and $\eta=0.05$. The results are shown in Fig. \ref{fig:coracon} noting that the shown values of GNMSE are averaged over $30$ repeated experiments. The GNMSE curve is not as smooth as in the previous case since fewer repeated experiments are carried out due to the larger network size and higher computational cost. 

For the Email-EU-Core dataset, the value of $\|\textbf{d}\Vert^2_{max}$ is found out to be $107$, resulting a theoretical value $\sigma_{ed}^2=15.29$. For simulations, we set $E=3$, $D=M=403$, and $\eta=0.05$. The results are shown in Fig. \ref{fig:eec}. The shown values of GNMSE are averaged over $50$ repeated experiments.

For the Wikipedia-Math-Daily dataset, the value of $\|\textbf{d}\Vert^2_{max}$ is found out to be $671$ followed by a theoretical value $\sigma_{ed}^2=95.67$. For simulations, we set $E=3$, $D=500$, and $\eta=0.03$. The comparison between the simulation result and the theoretical value is shown in Fig. \ref{fig:wikimath}. The shown values of GNMSE are averaged over $50$ repeated experiments.

\ifCLASSOPTIONonecolumn
\begin{figure}[!t]
    \centering
    \subfloat[]{\includegraphics[width=8cm]{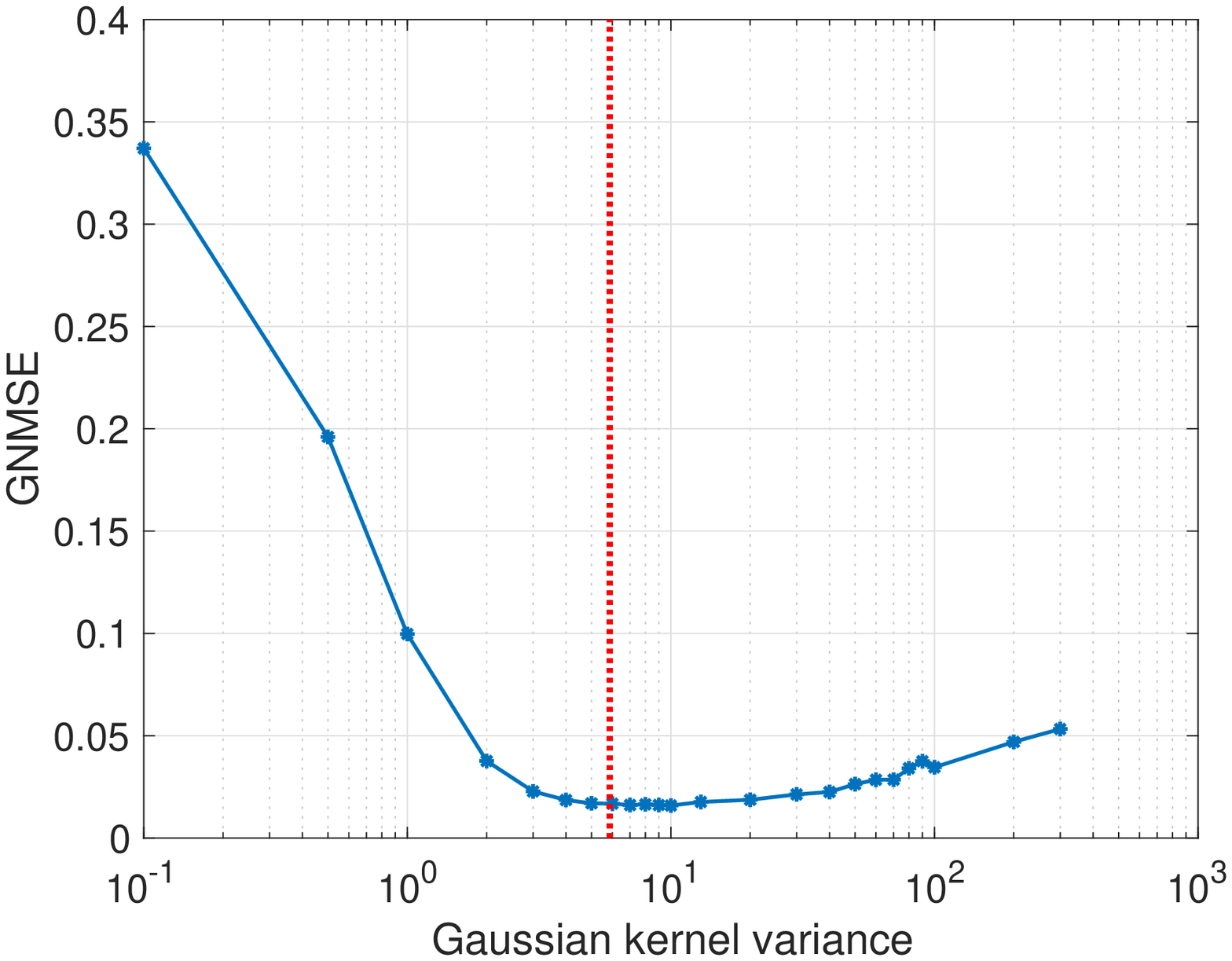}
    \label{fig:temperaturejan}}
    \subfloat[]{\includegraphics[width=8cm]{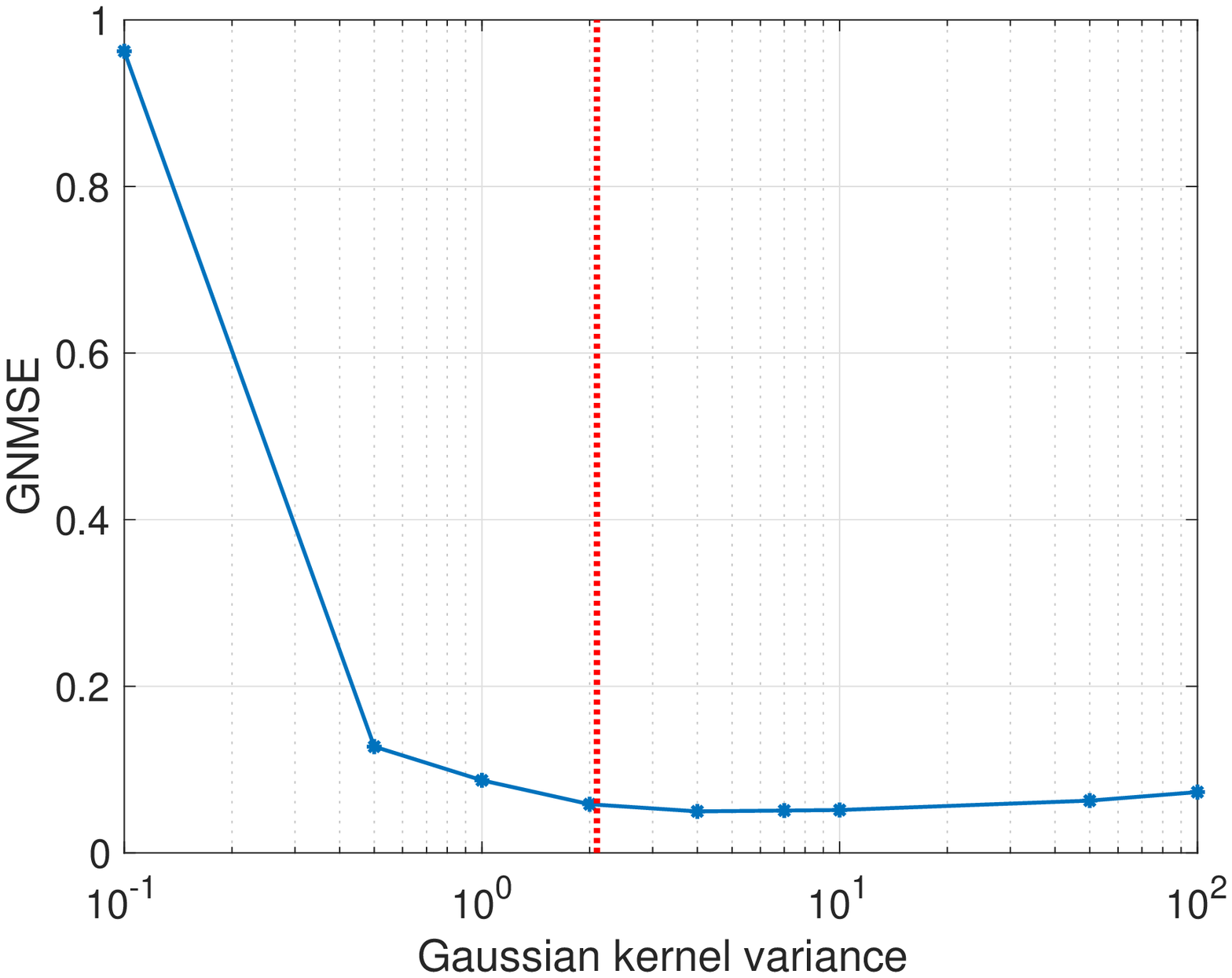}
    \label{fig:temperatureweighted}}
    \hfill
    \subfloat[]{\includegraphics[width=8cm]{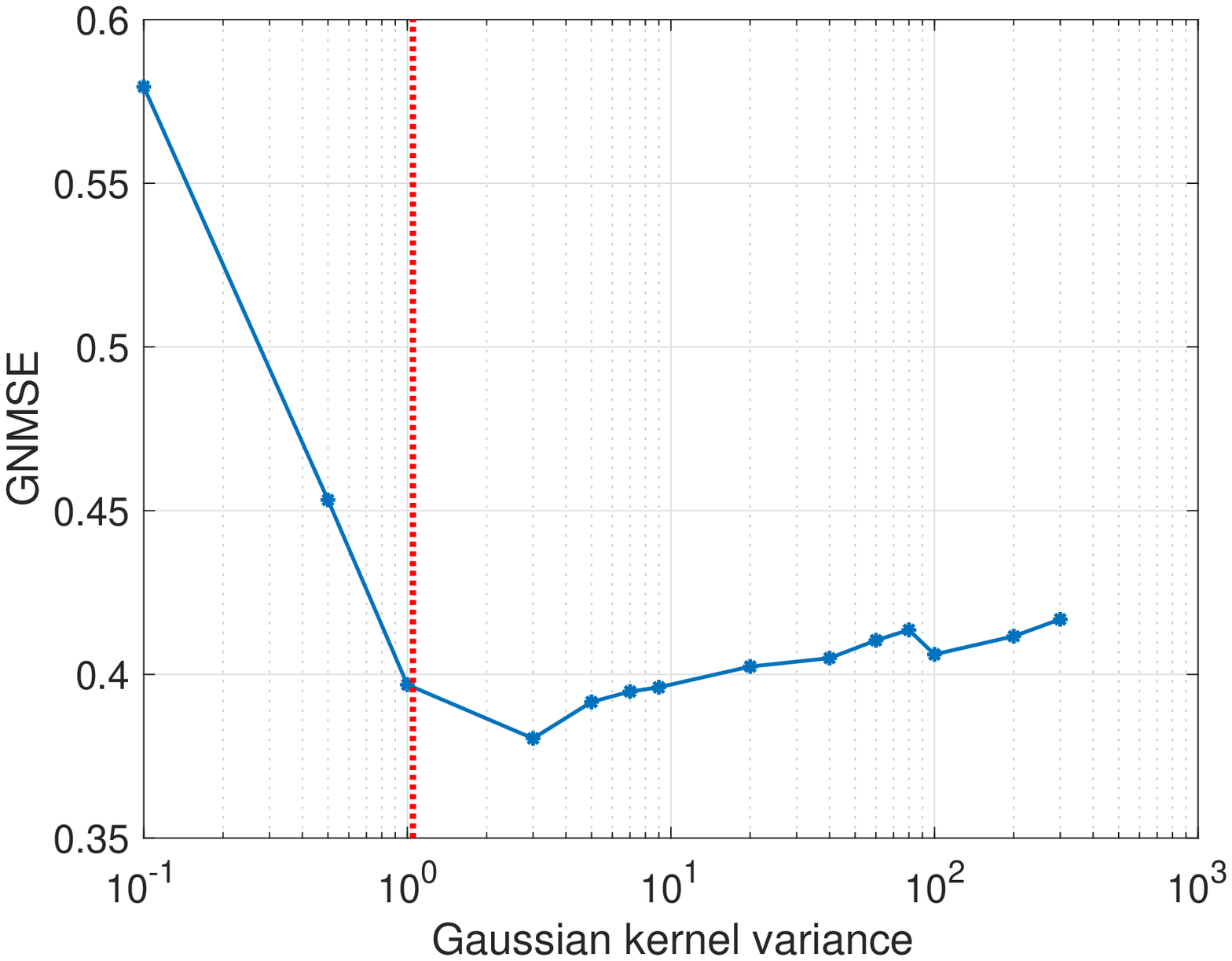}
    \label{fig:coracon}}
    \subfloat[]{\includegraphics[width=8cm]{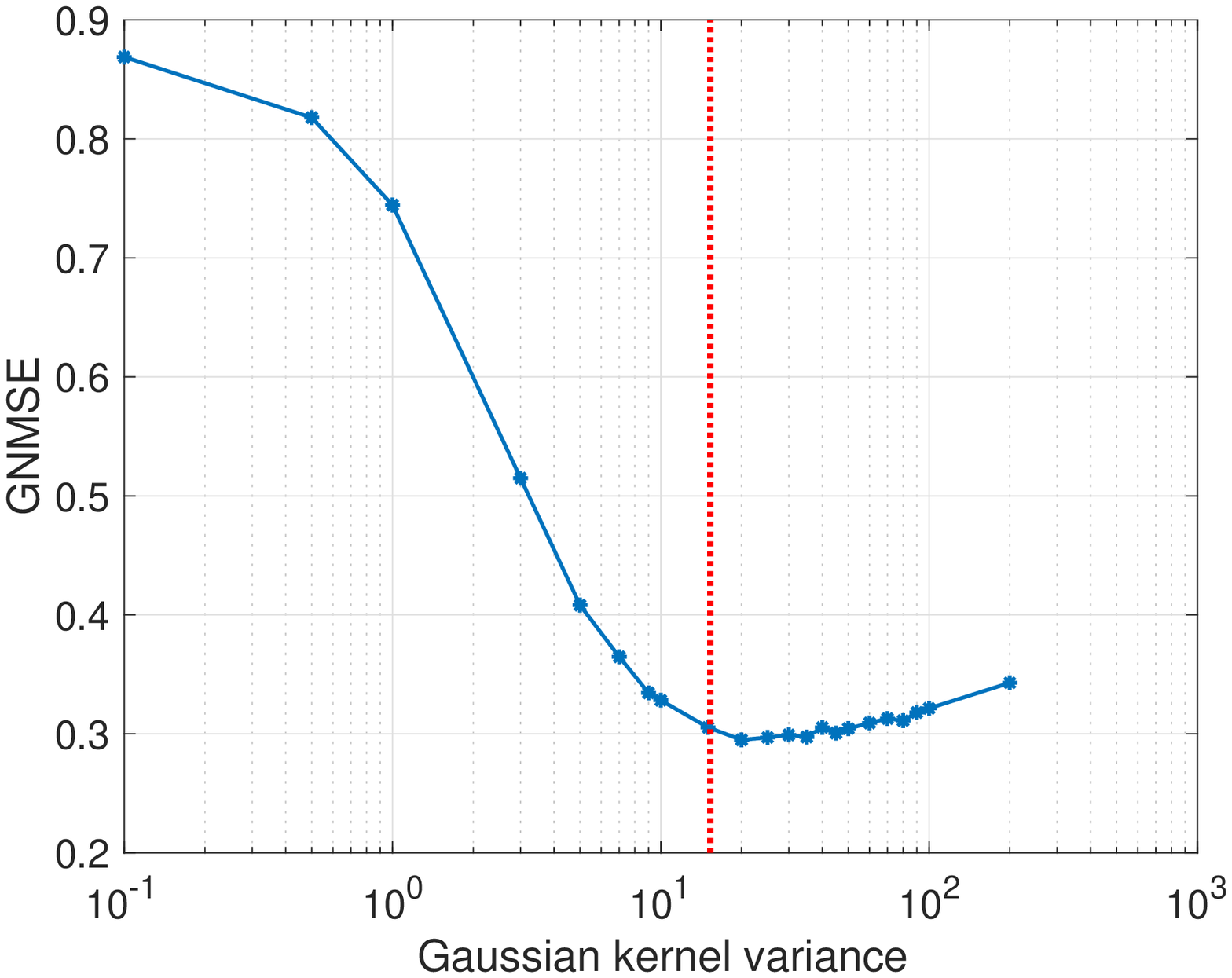}
    \label{fig:eec}}
    \hfill
    \subfloat[]{\includegraphics[width=8cm]{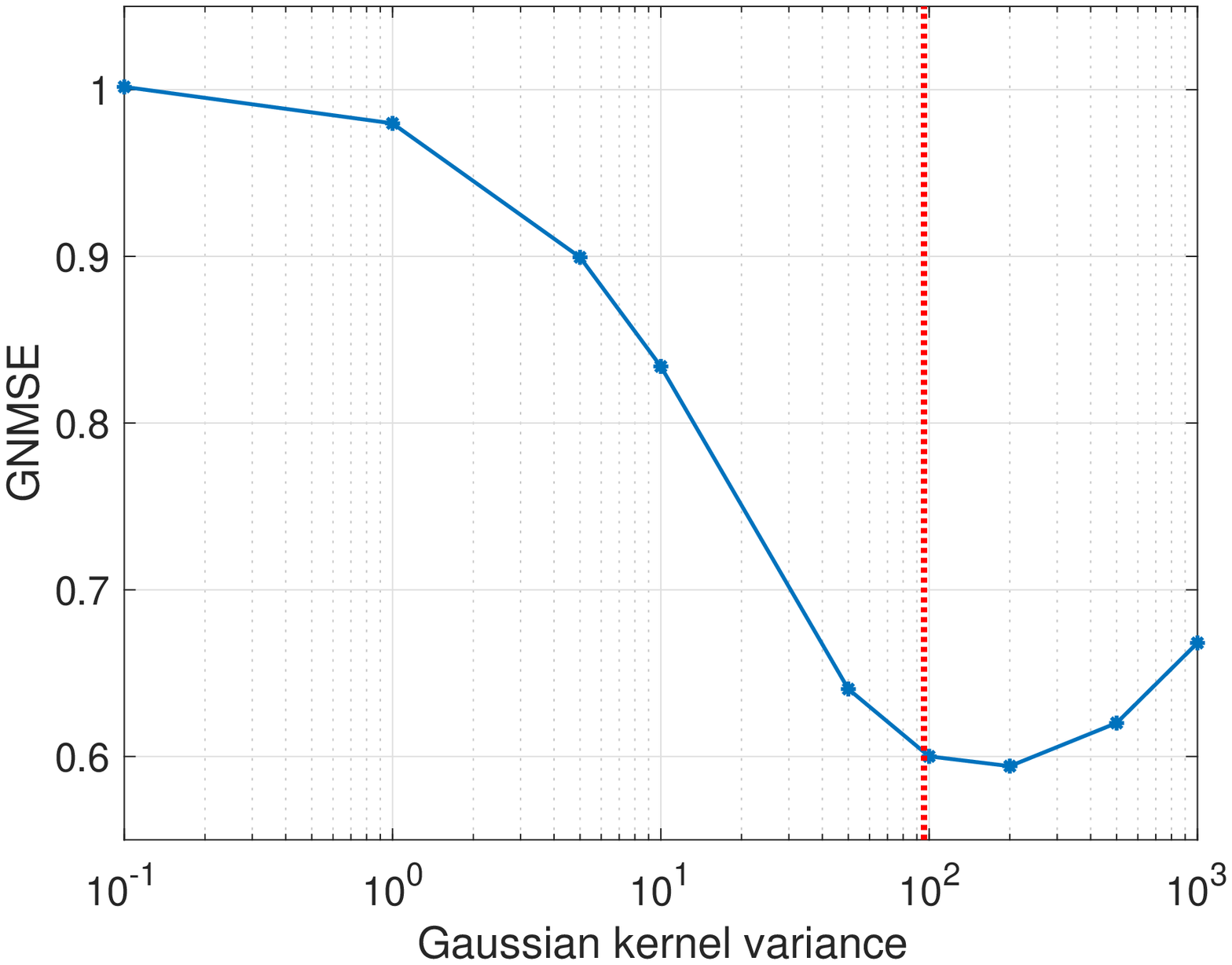}
    \label{fig:wikimath}}
    \caption{The GNMSE values with respect to the Gaussian kernel variance $\sigma^2$ for different datasets and graphs. (a) For the Temperature-Jan dataset with the unweighted graph. (b) For the Temperature-Jan dataset with the weighted graph. (c) For the Cora-Con dataset. (d) For the Email-EU-Core dataset. (e) For the Wiki-Math-Daily dataset.}
    \label{figs:gnmseVSsigma2}
\end{figure}
\else
\begin{figure}[!t]
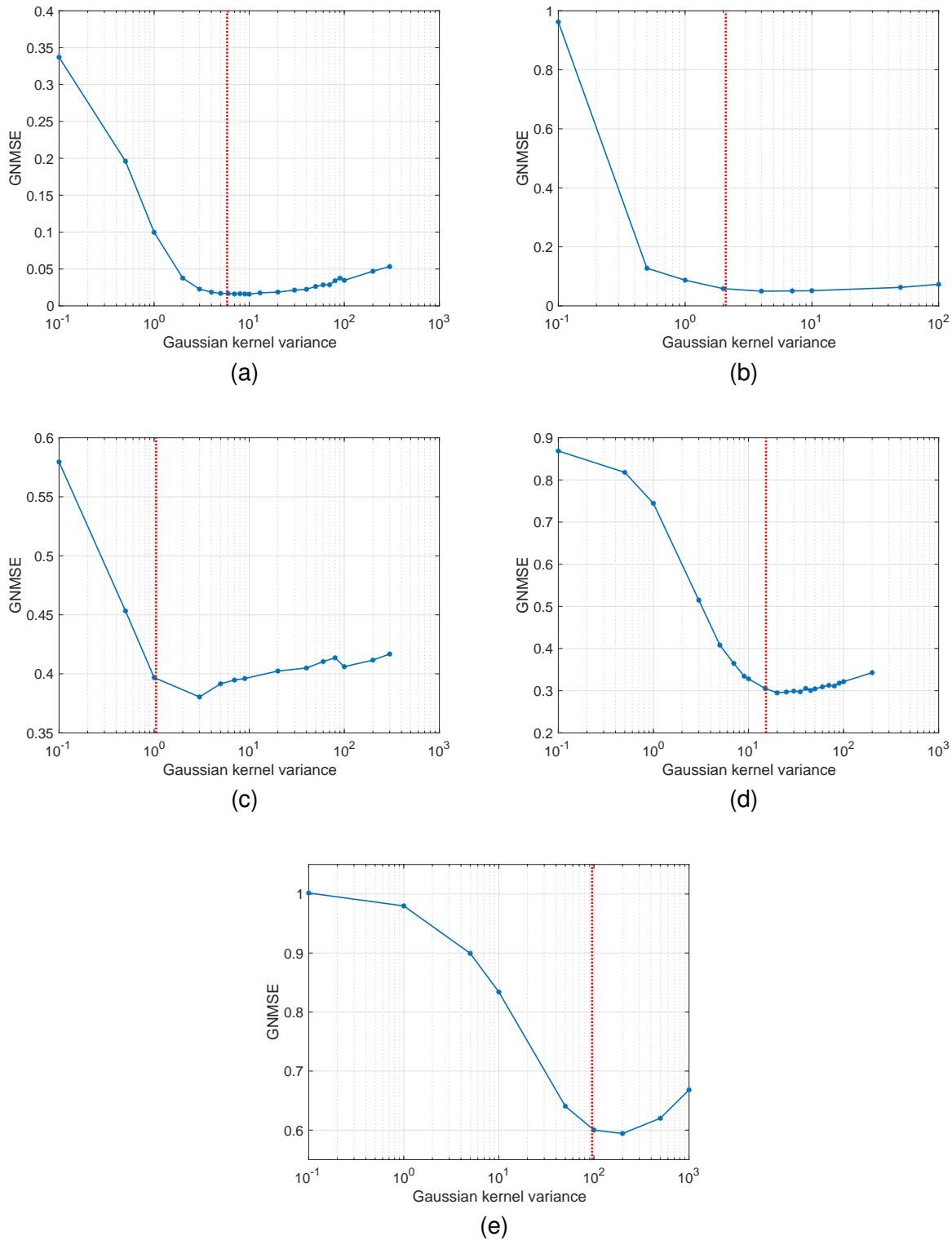

    \centering
    \subfloat[]{\includegraphics[width=4.5cm]{temperature-jan.eps}
    \label{fig:temperaturejan}}
    \subfloat[]{\includegraphics[width=4.5cm]{temperatureweighted.eps}
    \label{fig:temperatureweighted}}
    \hfill
    \subfloat[]{\includegraphics[width=4.5cm]{coracon.eps}
    \label{fig:coracon}}
    \subfloat[]{\includegraphics[width=4.5cm]{eec.eps}
    \label{fig:eec}}
    \hfill
    \subfloat[]{\includegraphics[width=4.5cm]{wikimathdaily.eps}
    \label{fig:wikimath}}
    \caption{The GNMSE values with respect to the Gaussian kernel variance $\sigma^2$ for different datasets and graphs. (a) For the Temperature-Jan dataset with the unweighted graph. (b) For the Temperature-Jan dataset with the weighted graph. (c) For the Cora-Con dataset. (d) For the Email-EU-Core dataset. (e) For the Wiki-Math-Daily dataset.}
    \label{figs:gnmseVSsigma2}
\end{figure}
\fi

\section{Discussions}
\label{sec:disc}

\subsection{Complexity of the Proposed Algorithm}
The majority of computational source for the proposed algorithm is used for calculating $\|\textbf{d}_{i,T+1}\Vert^2$ for all pairs of sampled nodes. For a training set containing $N$ sampled nodes, it would take $N(N-1)/2$ vector additions and inner products. Finding the largest $\|\textbf{d}_{i,T+1}\Vert^2$ value can be done along with its calculation, and takes minor computational resource and memory resource.

\subsection{Performance of the Proposed Algorithm}
\subsubsection{Impact From Dataset Statistics}
\label{subsubsec:algorithmperformance}
It is noticed that the proposed algorithm finds the best $\sigma^2$ in terms of GNMSE for the Temperature-Jan dataset, whereas for the other three datasets, the best $\sigma^2$ is slightly larger than what is found via the proposed algorithm. That is partly because similar adjacency vectors leading to similar nodal values is not guaranteed in later datasets. We take the comparison between the Temperature-Jan dataset and the Cora-Con dataset as an example. 

\begin{figure}[!t]
    \centering
    \includegraphics[width=8cm]{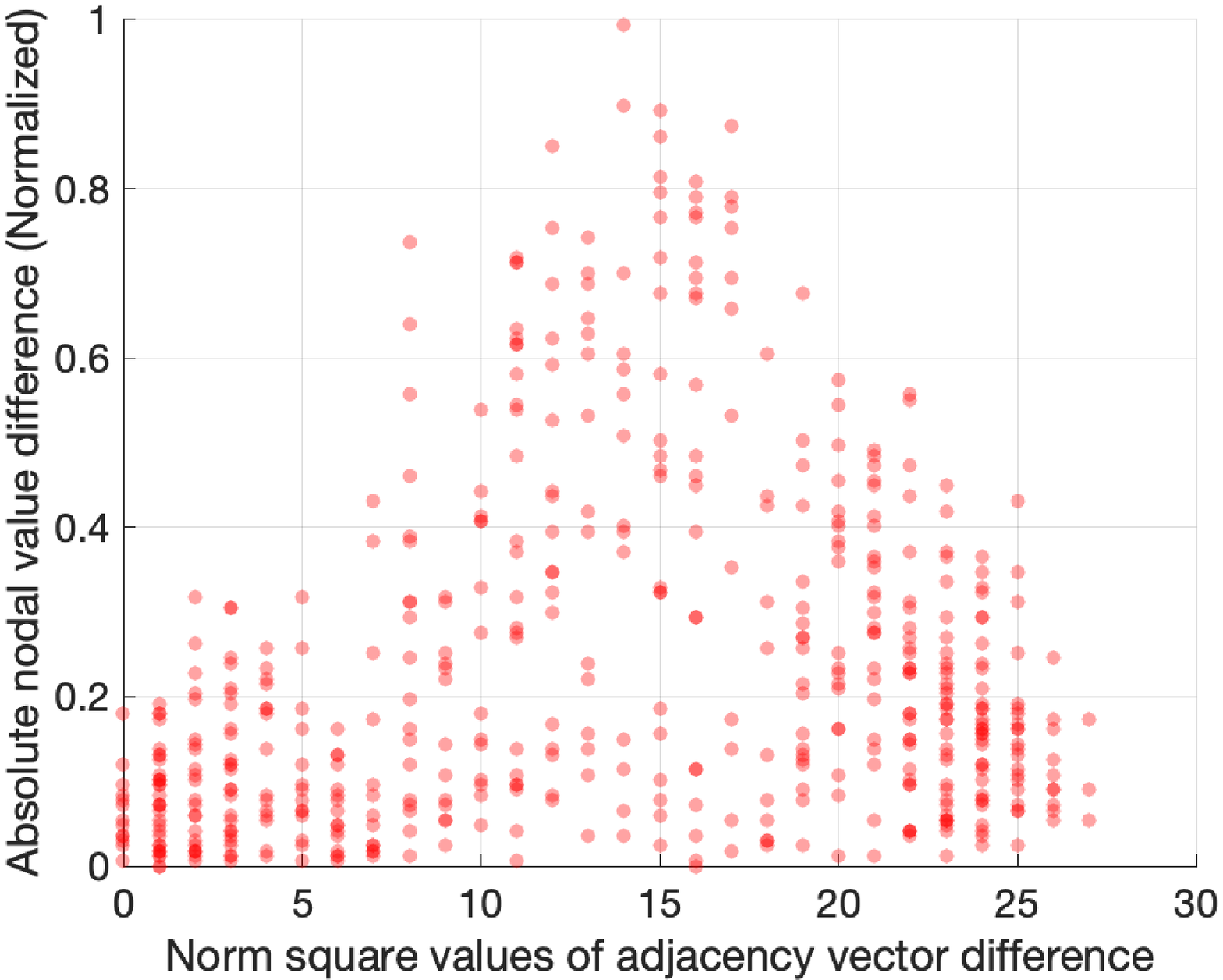}
    \caption{Scatter plot of the relationship between norm square of adjacency vector difference $\|\textbf{d}_{i,j}\Vert^2$ and absolute difference between nodal values $|y_i-y_j\vert$ for all pairs of sampled nodes in the Temperature-Jan dataset.}
    \label{fig:temperaturejanscatter}
\end{figure}

\begin{figure}[!t]
    \centering
    \includegraphics[width=8cm]{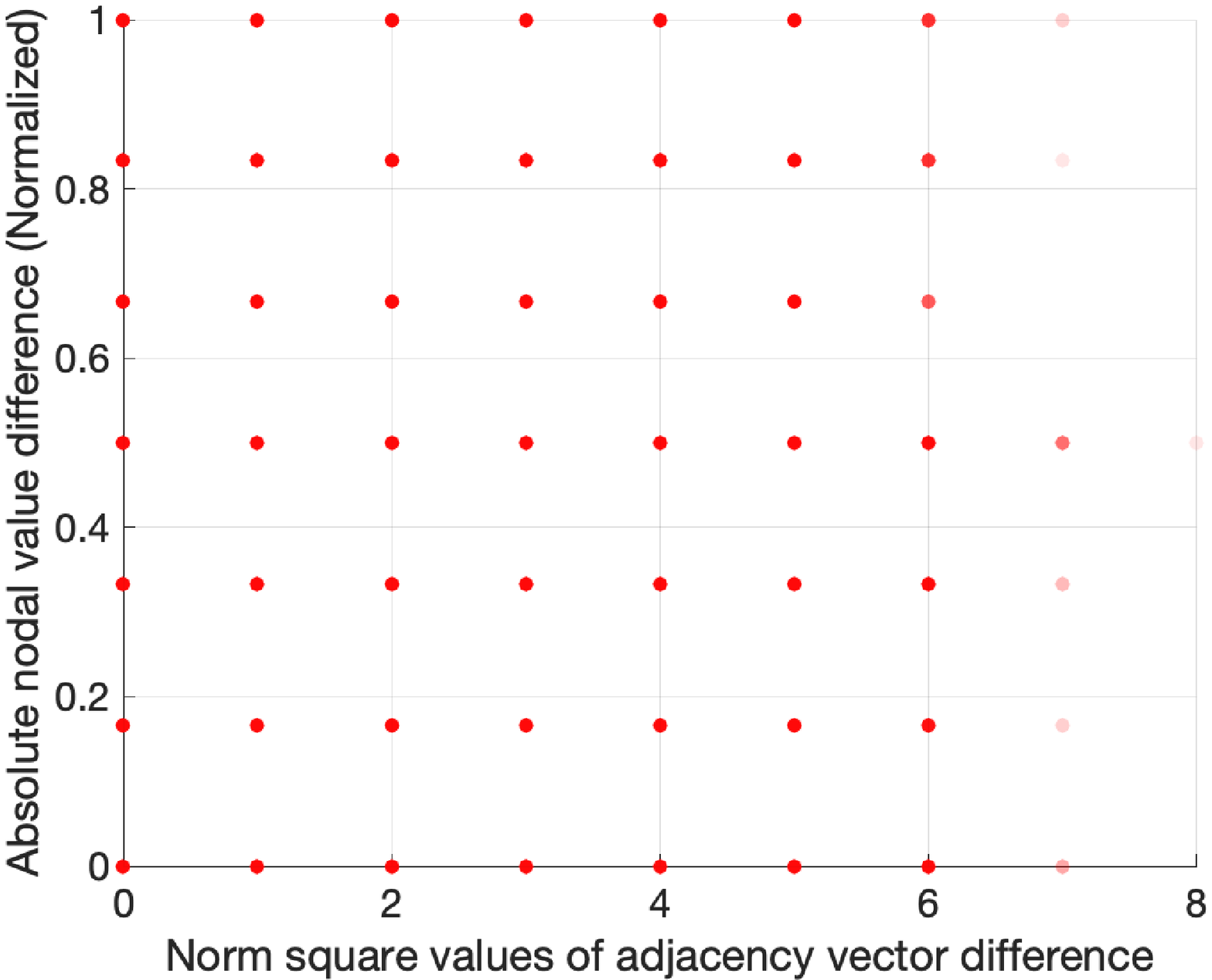}
    \caption{Scatter plot of the relationship between norm square of adjacency vector difference $\|\textbf{d}_{i,j}\Vert^2$ and absolute difference between nodal values $|y_i-y_j\vert$ for all pairs of sampled nodes in the Cora-Con dataset.}
    \label{fig:coraconscatter}
\end{figure}

Let us first understand the relationship between similarity in nodal values and similarity in adjacency vectors in the two datasets. We use $\|\textbf{d}_{i,j}\Vert^2$ to show dissimilarity of adjacency vectors and $|y_i-y_j\vert$ to show dissimilarity of nodal values for a pair of nodes. Notice the smaller the values, the more the similarity. 
Fig. \ref{fig:temperaturejanscatter} and Fig. \ref{fig:coraconscatter} show $|y_i-y_j\vert$ versus $\|\textbf{d}_{i,j}\Vert^2$ for the two datasets, respectively.

In Fig. \ref{fig:temperaturejanscatter}, the horizontal axis takes discrete values because the graph is unweighted, and the vertical axis takes continuous values because the nodal values, i.e., temperature for the stations, take continuous values. More importantly, it is seen that when $\|\textbf{d}_{i,j}\Vert^2$ is small, the nodal value difference is also small. For instance, when $\|\textbf{d}_{i,j}\Vert^2=5$ for a pair of nodes, $|y_i-y_j\vert$ may take a value within $[0,0.35]$. Whereas, for a pair of nodes with $\|\textbf{d}_{i,j}\Vert^2=15$, the nodal value difference is likely to be greater than $0.3$ and could be up to $1$. (The behavior for $\|\textbf{d}_{i,j}\Vert^2\geq15$ corresponds to the fact that stations with higher altitudes have smaller temperature difference. Although such stations may largely vary in altitudes and thus adjacency vectors, they have relatively low temperature.) This property of the dataset is due to the fact that connected nodes have closer altitudes. Since connected nodes tend to have similar adjacency vectors in the graph, considering correlation between altitudes and temperature, it is expected that similar adjacency vectors tend to have closer nodal values (temperature).

In Fig. \ref{fig:coraconscatter}, the horizontal axis takes discrete values. The vertical axis also takes discrete values because the nodal labels in the dataset represent categories. It is seen that $|y_i-y_j\vert$ takes large values with nonnegligible probability even when $\|\textbf{d}_{i,j}\Vert^2=0$. Recalling that the graph indicates citing behavior among papers, such a relationship between nodal values and adjacency vectors implies that the category of a citing paper is weakly correlated with the categories of the cited papers. That is, it is possible that two papers citing the same reference belong to different classes, and that two papers with different citing patterns are categorically the same. From the SKG model perspective, similarity in adjacency vectors may not lead to similarity in nodal values in the Cora-Con dataset. 

Now we explain how such a relation between adjacency vectors and nodal values impacts the choice of best $\sigma^2$. Recall at the boundary between the Extending Range and the Disturbing Range, the efficient distance shown in Fig. \ref{fig:distance} divides sampled nodes based on similarity in adjacency vectors, and the prediction can be considered as a weighted average of the inner nodes. For the Temperature-Jan dataset, the inner nodes have small nodal value difference from the tested node, and thus the prediction would be close. Whereas, for the Cora-Con dataset, the predictions are precise for some tested nodes while they have large errors for the rest. However, these large errors can be reduced in the Disturbing Range. Recall at this range, predictions are weighted averages over all sampled nodes. Large errors at the boundary $\sigma_{ed}^2$ will become less in the Disturbing Range, but meanwhile, small errors may increase. Since GNMSE penalizes more on larger error, GNMSE performance could get better in the Disturbing Range than at $\sigma_{ed}^2$ for the Cora-Con dataset.

In summary, for datasets where similar adjacency vectors lead to dissimilar nodal values with nonnegligible probability, the resulting $\sigma_{ed}^2$ of the proposed algorithm is smaller than the best $\sigma^2$ which is in the Disturbing Range in terms of GNMSE.

\subsubsection{Impact From Other Hyperparameters}
We would like to note that our analysis applies for ideal cases where the number of random feature $D$ is sufficiently large and the learning rate $\eta$ and the number of epochs $E$ are properly chosen.

If $D$ is not large enough, the validity of the exponential approximation of $B_{i,j}$ would weaken, leading to larger GNMSE values. The impact from $D$ applies to all $\sigma^2$ such that the chosen $D$ does not affect the best $\sigma^2$ theoretically. Our proposed algorithm provides $\sigma^2$ which is near-optimal, $\sigma^2_{ed}$ is affected by $D$ without \textbf{Algorithm \ref{alg:findnoiserange}}. However, the impact of $D$ decays as $D$ increases due to the $\log(\cdot)$ function in the denominator of (\ref{equ:edthresh}). In practice, $D$ is usually more than tens or hundreds and thus the impact of $D$ on $\sigma^2_{ed}$ is limited.

The parameters $\eta$ and $E$ should be jointly chosen such that $\sum_{i=1}^{EN}F_{i,EN+1}\approx1$. Note that $\eta$ cannot be very large to avoid parameter explosion. When $\eta$ becomes smaller, $E$ should be increased accordingly. In this case, our analysis applies and $\sigma^2_{ed}$ is close to optimum. If $E$ is not sufficient either, $\sigma^2_{ed}$ would not be ideal because the weighting property of $F_{i,j}$ is invalid. In this case, the best $\sigma^2$ should be much larger than $\sigma^2_{ed}$ to mitigate the biased sum of the contribution weights.

\subsection{Similarity Transfer of SKG}
We refer the reader to Figs. \ref{fig:temperaturejan}--\ref{fig:eec}. Note that GNMSE values vary considerably among the three datasets. Thus, it is believed that SKG has better performance on some datasets than others. This can be understood via the weighting property together with the characteristic of a dataset. With $\sigma_{ed}^2$, predictions are weighted averages of a group of nodes. The weights are acquired from adjacency vectors and applied for nodal values. If similar adjacency vectors do not lead to similar nodal values, predictions would have large errors. If similar adjacency vectors lead to dissimilar nodal values with nonnegligible probability in a dataset, the SKG performance would be less ideal on the dataset. This is clear when comparing SKG performance between Temperature-Jan dataset on which GNMSE could be lower than $0.05$ and the Cora-Con dataset on which GNMSE values are above $0.35$. From the comparison, we concluded that SKG with a Gaussian kernel assumes that similarity of adjacency vectors leads to similarity of nodal values. 

\subsection{Extension of the Analysis}
We used SKG with a Gaussian kernel to illustrate our analysis framework based on similarity measures and contribution weights. The analysis framework is not constrained in the cases of Gaussian kernels. In fact, it is applicable to all shift-invariant kernels \cite{RandomFeatures}. Specifically speaking, as we mentioned before, the exponential approximation is the same expression as the random feature approximation but in a reverse order. So, any shift-invariant kernel that has a Fourier transform can easily build its similarity measure. In what follows, the proposed properties of contribution weights apply no matter what kernel is used because they are built only on the similarity measure. Take Laplacian kernels as an example. The mathematical expression of a Laplacian kernel is $\kappa_l(\textbf{a}_i,\textbf{a}_j)=e^{-\frac{\|\textbf{a}_i-\textbf{a}_j\Vert}{b}}$ where $\textbf{a}_i$ and $\textbf{a}_j$ are two input vectors, $\|\cdot\Vert$ denotes the $l_1$-norm, and $b$ is the tunable diversity. Then we can build $B_{i,j}\cong2\eta e^{-\frac{\|\textbf{a}_i-\textbf{a}_j\Vert}{b}}$ with a sufficiently large $D$. As a result, we have the requirement $2\eta e^{-\frac{\|\textbf{d}\Vert_{max}}{b}}=noise_{up}$ where $\|\textbf{d}\Vert_{max}$ denotes the maximum $l_1$-norm among all pairs of sampled nodes to find a suitable diversity $b$.

Our analysis framework can be helpful for other algorithms which have a (shift-invariant-)kernel-based learning model. For example, GKLMS-RFF in \cite{GradrakerTimeInfo} has the same learning model, but uses filtered nodal value time series by a known graph filter as the model input. Our analysis can explicitly point out how to configure the model in the GKLMS-RFF algorithm. In addition, our analysis indicates a requirement on the graph filter that similarity in the filtered nodal value time series should result in similarity of the nodal value, since the model assumes that similarity of inputs leads to similarity of outputs.

\section{Conclusions}
\label{sec:conclu}

The paper dealt with the problem of how to choose a suitable Gaussian kernel for SKG given a training set. Two variables, the similarity measure $B_{i,j}$ and the contribution weight $F_{i,j}$ as well as their properties were introduced. Using the properties, we were able to find the impact of Gaussian kernel variance in SKG and divide possible $\sigma^2$ range into four ranges, i.e., Chaos Range, Extending Range, Disturbing Range, and Averaging Range. Given detailed $B_{i,T+1}$ and $F_{i,T+1}$ behavior in each range, we conclude that the boundary between Extending Range and Disturbing Range should be a suitable $\sigma^2$ for SKG. Important properties of the introduced variables have been confirmed by simulations. Effectiveness of the proposed algorithm on the Temperature-Jan dataset, the Cora-Con dataset, the Email-EU-Core dataset, and the Wikipedia-Math-Daily has been verified in experiments. Additionally, by comparison of GNMSE between the Temperature-Jan dataset and the Cora-Con dataset, we drew a conclusion that SKG assumes that similarity of adjacency vectors leads to similarity of nodal values. 

\section{Acknowledgements}
\label{sec:ack}

We thank the anonymous reviewers whose comments improved the quality of the paper.

\appendix[Proofs of Claims]

\textbf{Proof of Claim 1}. Since the value of a Gaussian kernel relates with the difference of its input vectors, its expression can also be $\kappa(\textbf{x})$ $\kappa(\textbf{x})=e^{-\frac{\|\textbf{x}\Vert^2}{2\sigma^2}}$ where $\textbf{x}=[x_1,...,x_N]^\top$, and its Fourier transform $\rho_\textbf{f}(\textbf{f})$ can be written as in \cite{nDimFourier}
\begin{equation*}
\rho_\textbf{f}(\textbf{f})=\int_{\textbf{R}^n}\kappa(\textbf{x})e^{-j2\pi\textbf{f}^\top\textbf{x}}d\textbf{x}
\end{equation*}
where $\textbf{f}=[f_1,f_2,...,f_N]^\top$, and $\textbf{f}^\top\textbf{x}$ denotes the inner product of $\textbf{f}$ and $\textbf{x}$, i.e., $\textbf{f}^\top\textbf{x}=\sum_{i=1}^Nf_ix_i$. Since $\textbf{x}\in\textbf{R}^n$, $\|\textbf{x}\Vert^2=\sum_{i=1}^Nx_i^2$, then
\begin{align*}
\rho_\textbf{f}(\textbf{f})&=\int_{-\infty}^\infty...\int_{-\infty}^\infty e^{-\frac{\sum_{i=1}^Nx_i^2}{2\sigma^2}}e^{-j2\pi\sum_{i=1}^Nf_ix_i}dx_1...dx_N\\
&=\prod_{i=1}^N\int_{-\infty}^\infty e^{-\frac{x_i^2}{2\sigma^2}}e^{-j2\pi f_ix_i}dx_i
\end{align*}
which can be seen as the product of the Fourier transform for the one-dimensional Gaussian kernels in every dimension. Notice that the product implies independence among dimensions.

It is a well-known Fourier transform pair that
\begin{equation*}
e^{-\pi x^2}\stackrel{\mathcal{F.\,T.}}{\longleftrightarrow}e^{-\pi f^2}.
\end{equation*}
Notice the Fourier transform of $g(x)$, $\mathcal{F}\{g(x)\}=G(f)=\int_{-\infty}^\infty g(x)e^{-j2\pi fx}dx$. According to the time scaling property of Fourier transform, i.e., $g(ax)\stackrel{\mathcal{F.\,T.}}{\longleftrightarrow}\frac{1}{|a\vert}G(\frac{f}{a})$ where $g(x)\stackrel{\mathcal{F.\,T.}}{\longleftrightarrow}G(f)$, we have
\begin{equation*}
e^{-\frac{x^2}{2\sigma^2}}\stackrel{\mathcal{F.\,T.}}{\longleftrightarrow}\sqrt{2\pi}\sigma e^{-2\pi^2\sigma^2f^2}=\frac{1}{(\frac{1}{2\pi\sigma})\sqrt{2\pi}}e^{-\frac{f^2}{2(\frac{1}{2\pi\sigma})^2}}.
\end{equation*}
That is, $f\thicksim\mathcal{N}(0,(\frac{1}{2\pi\sigma})^2)$. For $\xi=2\pi f$, it is $\xi\thicksim\mathcal{N}(0,\sigma^{-2})$. Because when $X$ is a random variable and $Y=aX+b$ where $a$ and $b$ are both constants, the probability density function (PDF) of $Y$ is $p_Y(y)=\frac{1}{|a\vert}p_X(\frac{y-b}{a})$ where $p_X(x)$ denotes the PDF of $X$.

Similarly, all elements in $\boldsymbol{\xi}$ have Gaussian distribution $\mathcal{N}(0,\sigma^{-2})$, that is, $\boldsymbol{\xi}\thicksim\mathcal{N}(0,\sigma^{-2}\textbf{I}_N)$.\hfill $\blacksquare$\\

\textbf{Proof of Claim 2}. The characteristic function \cite{CharacteristicFunctions} of the random variable $X$ is
\begin{equation*}
\E[e^{j\upsilon X}]\equiv\psi(j\upsilon)=\int_{-\infty}^\infty e^{j\upsilon x}p(x)dx=\mathcal{F}\{p(x)\}
\end{equation*}
where $\upsilon$ is real, $j=\sqrt{-1}$, $\mathcal{F}$ denotes Fourier transform, and $p(x)=\frac{1}{\sigma_X\sqrt{2\pi}}e^{-\frac{x^2}{2\sigma_X^2}}$. According to \textrm{\textbf{Claim 1}}, it is clear that
\begin{equation*}
p(x)=\frac{1}{\sigma_X\sqrt{2\pi}}e^{-\frac{x^2}{2\sigma_X^2}}\stackrel{\mathcal{F.T.}}{\longleftrightarrow}\psi(j\upsilon)=e^{-\frac{\sigma_X^2\upsilon^2}{2}}.
\end{equation*}
Thus,
\begin{equation*}
\E[e^{jX}]=\psi(j\upsilon)|_{\upsilon=1}=e^{-\frac{\sigma_X^2}{2}}.
\end{equation*}
On the other hand, according to Euler's Equation, $e^{jX}=\cos(X)+j\sin(X)$, and

\ifCLASSOPTIONonecolumn
\begin{equation*}
\E[e^{jX}]=\E[\cos(X)+j\sin(X)]=\E[\cos(X)]+j\E[\sin(X)].
\end{equation*}
\else
\begin{align*}
\E[e^{jX}]=&\E[\cos(X)+j\sin(X)]\\
=&\E[\cos(X)]+j\E[\sin(X)].
\end{align*}
\fi

Since Gaussian PDFs with zero mean is an even function while $\sin(\cdot)$ is odd, $\E[\sin(X)]=0$,
\begin{equation*}
\E[\cos(X)]=\E[e^{jX}]=e^{-\frac{\sigma_X^2}{2}}.
\end{equation*}
Similarly,
\begin{equation*}
\E[\cos(2X)]=\E[e^{j2X}]=\psi(j\upsilon)|_{\upsilon=2}=e^{-2\sigma_X^2}.
\end{equation*}
Then,
\begin{align*}
\Var[e^{jX}]=&\E\left[\cos^2(X)\right]-\left(\E\left[\cos(X)\right]\right)^2\\
=&\frac{1}{2}\E\left[\cos(2X)\right]+\frac{1}{2}-\left(\E\left[\cos(X)\right]\right)^2\\
=&\frac{1}{2}\left(e^{-\sigma_X^2}-1\right)^2
\end{align*}
$\hfill\blacksquare$\\

\textbf{Proof of Claim 3}. First we prove for the trainable parameter $\boldsymbol{\theta}$ at time $t$ that
\begin{equation*}
\boldsymbol{\theta}_t=2\eta\sum_{i=1}^t{e_iz(\textbf{a}_i)}
\end{equation*}
where $e_i$ is the predicting error at time $i$, $e_i=y_i-\hat{f}_i$. Note that the square loss is $\mathcal{L}(y_{pred},y_{true})=(y_{true}-y_{pred})^2$ in our case, the loss at time $t$ is
\begin{equation*}
\mathcal{L}(y_t,\boldsymbol{\theta}_{t-1}^\top z(\textbf{a}_t))=(f_t-\boldsymbol{\theta}_{t-1}^\top z(\textbf{a}_t))^2.
\end{equation*}
Then the gradient of the loss function with respect to $\boldsymbol{\theta}_{t-1}$, i.e., $g_t$, is
\begin{align*}
g_t&=\nabla_{\boldsymbol{\theta}_{t-1}}\mathcal{L}(y_t,\boldsymbol{\theta}_{t-1}^\top z(\textbf{a}_t))\\
&=-2(f_t-\boldsymbol{\theta}_{t-1}^\top z(\textbf{a}_t))z(\textbf{a}_t)=-2e_tz(\textbf{a}_t)
\end{align*}
which uses
\ifCLASSOPTIONonecolumn
\begin{align*}
\nabla_\textbf{x}(\textbf{x}^\top\textbf{a})&=\nabla_\textbf{x}(\textbf{a}^\top\textbf{x})=[\frac{\partial(\textbf{a}^\top\textbf{x})}{\partial x_1},\frac{\partial(\textbf{a}^\top\textbf{x})}{\partial x_2},...,\frac{\partial(\textbf{a}^\top\textbf{x})}{\partial x_N}]^\top\\
&=[a_1,a_2,...,a_N]^\top=\textbf{a}
\end{align*}
\else
\begin{align*}
\nabla_\textbf{x}(\textbf{x}^\top\textbf{a})&=\nabla_\textbf{x}(\textbf{a}^\top\textbf{x})\\
&=[\frac{\partial(\textbf{a}^\top\textbf{x})}{\partial x_1},\frac{\partial(\textbf{a}^\top\textbf{x})}{\partial x_2},...,\frac{\partial(\textbf{a}^\top\textbf{x})}{\partial x_N}]^\top\\
&=[a_1,a_2,...,a_N]^\top=\textbf{a}
\end{align*}
\fi
where $\textbf{a}=[a_1,a_2,...,a_N]^\top$ and $\textbf{x}=[x_1,x_2,...,x_N]^\top$. Denote $\eta$ as the learning rate, then
\begin{equation*}
\boldsymbol{\theta}_t=\boldsymbol{\theta}_{t-1}-\eta g_t=\boldsymbol{\theta}_{t-1}+2\eta e_tz(\textbf{a}_t).
\end{equation*}
Tracing back to $\boldsymbol{\theta}_0$, $\boldsymbol{\theta}_t$ can be rewritten in terms of $\boldsymbol{\theta}_0$ as
\begin{align*}
\boldsymbol{\theta}_t&=\boldsymbol{\theta}_{t-1}+2\eta e_tz(\textbf{a}_t)\\
&=\boldsymbol{\theta}_{t-2}+2\eta e_{t-1}z(\textbf{a}_{t-1})+2\eta e_tz(\textbf{a}_t)\\
&=\boldsymbol{\theta}_0+2\eta\sum_{i=1}^t[e_iz(\textbf{a}_i)].
\end{align*}
Together with the initialization $\boldsymbol{\theta}_0=\textbf{0}$, we get
\begin{equation*}
\boldsymbol{\theta}_t=2\eta\sum_{i=1}^t[e_iz(\textbf{a}_i)].
\end{equation*}
Then the prediction at time $t$, $\hat{f}_t$, can be expressed as
\begin{equation*}
\hat{f}_t=\boldsymbol{\theta}_{t-1}^\top z(\textbf{a}_t)=2\eta\sum_{i=1}^{t-1}[e_iz^\top(\textbf{a}_i)z(\textbf{a}_t)]=\sum_{i=1}^{t-1}{e_iB_{i,t}}
\end{equation*}
which uses the definition of $B_{i,j}$ in (\ref{equ:bijdef}). Now we are prepared to prove the claimed equality. First, check the claimed equality for $t=2,3,4$. It should be
\ifCLASSOPTIONonecolumn
\begin{align*}
\hat{f}_2=&y_1B_{1,2}=y_1F_{1,2}\\
\hat{f}_3=&y_2B_{2,3}+y_1(B_{1,3}-B_{1,2}B_{2,3})=y_2F_{2,3}+y_1F_{1,3}\\
\hat{f}_4=&y_3B_{3,4}+y_2(B_{2,4}-B_{2,3}B_{2,4})+y_1[B_{1,4}-B_{1,3}B_{3,4}-B_{1,2}(B_{2,4}-B_{2,3}B_{3,4})]\\
=&y_3F_{3,4}+y_2F_{2,4}+y_1F_{1,4}
\end{align*}
\else
\begin{align*}
\hat{f}_2=&y_1B_{1,2}=y_1F_{1,2}\\
\hat{f}_3=&y_2B_{2,3}+y_1(B_{1,3}-B_{1,2}B_{2,3})=y_2F_{2,3}+y_1F_{1,3}\\
\hat{f}_4=&y_3B_{3,4}+y_2(B_{2,4}-B_{2,3}B_{2,4})\\
&+y_1[B_{1,4}-B_{1,3}B_{3,4}-B_{1,2}(B_{2,4}-B_{2,3}B_{3,4})]\\
=&y_3F_{3,4}+y_2F_{2,4}+y_1F_{1,4}
\end{align*}
\fi
from which the equality holds. Suppose there exists $t\geq 4$ such that
\begin{equation*}
\hat{f}_k=\sum_{i=1}^{k-1}y_{i}F_{i,k}
\end{equation*}
where
\begin{equation*}
F_{i,k}=
	\begin{cases}
	B_{i,k}, & \text{ if } i=k-1,\\
	B_{i,k}-\sum_{j=i+1}^{k-1}B_{i,j}F_{j,k}, & \text{ if } 1\leq i<k-1.
	\end{cases}
\end{equation*}
holds for all integers $k\in\{2,3,...,t\}$. Then,
\ifCLASSOPTIONonecolumn
\begin{equation*}
\hat{f}_{t+1}=\sum_{i=1}^{t}{e_iB_{i,t+1}}=\sum_{i=1}^t{(y_i-\hat{f}_i)B_{i,t+1}}.
\end{equation*}
\else
\begin{align*}
\hat{f}_{t+1}&=\sum_{i=1}^{t}{e_iB_{i,t+1}}\\
&=\sum_{i=1}^t{(y_i-\hat{f}_i)B_{i,t+1}}.
\end{align*}
\fi
For $\hat{f}_1=0$, we have
\ifCLASSOPTIONonecolumn
\begin{align*}
\hat{f}_{t+1}&=\sum_{i=1}^t{y_iB_{i,t+1}}-\sum_{i=2}^t{\hat{f}_iB_{i,t+1}}\\
&=\sum_{i=1}^t{y_iB_{i,t+1}}-\sum_{i=2}^t\sum_{j=1}^{i-1}{y_jF_{j,i}B_{i,t+1}}=\sum_{i=1}^t{y_iB_{i,t+1}}-\sum_{j=1}^{t-1}\sum_{i=j+1}^t{y_jF_{j,i}B_{i,t+1}}\\
&=y_tB_{t,t+1}+\sum_{i=1}^{t-1}y_i(B_{i,t+1}-\sum_{j=i+1}^tF_{i,j}B_{j,t+1})\\
&=y_tB_{t,t+1}+\sum_{i=1}^{t-1}y_i\tilde{F}_{i,t+1}
\end{align*}
\else
\begin{align*}
\hat{f}_{t+1}&=\sum_{i=1}^t{y_iB_{i,t+1}}-\sum_{i=2}^t{\hat{f}_iB_{i,t+1}}\\
&=\sum_{i=1}^t{y_iB_{i,t+1}}-\sum_{i=2}^t\sum_{j=1}^{i-1}{y_jF_{j,i}B_{i,t+1}}\\
&=\sum_{i=1}^t{y_iB_{i,t+1}}-\sum_{j=1}^{t-1}\sum_{i=j+1}^t{y_jF_{j,i}B_{i,t+1}}\\
&=y_tB_{t,t+1}+\sum_{i=1}^{t-1}y_i(B_{i,t+1}-\sum_{j=i+1}^tF_{i,j}B_{j,t+1})\\
&=y_tB_{t,t+1}+\sum_{i=1}^{t-1}y_i\tilde{F}_{i,t+1}
\end{align*}
\fi
which shows the coefficient of $y_t$ here is $B_{t,t+1}$. Since the coefficient of $y_t$ in calculating $\hat{f}_{t+1}$ is $F_{t,t+1}$ according to definition of $F_{i,j}$ in (\ref{equ:fijdef}), the last equality justifies $F_{t,t+1}=B_{t,t+1}$. Besides, the coefficient for $y_{t-1}$ is
\ifCLASSOPTIONonecolumn
\begin{equation*}
\tilde{F}_{t-1,t+1}=B_{t-1,t+1}-F_{t-1,t}B_{t,t+1}=B_{t-1,t+1}-B_{t-1,t}F_{t,t+1}=F_{t-1,t+1}.
\end{equation*}
\else
\begin{align*}
\tilde{F}_{t-1,t+1}&=B_{t-1,t+1}-F_{t-1,t}B_{t,t+1}\\
&=B_{t-1,t+1}-B_{t-1,t}F_{t,t+1}=F_{t-1,t+1}.
\end{align*}
\fi
\begin{figure}[!t]
\centering
\includegraphics[width=8cm]{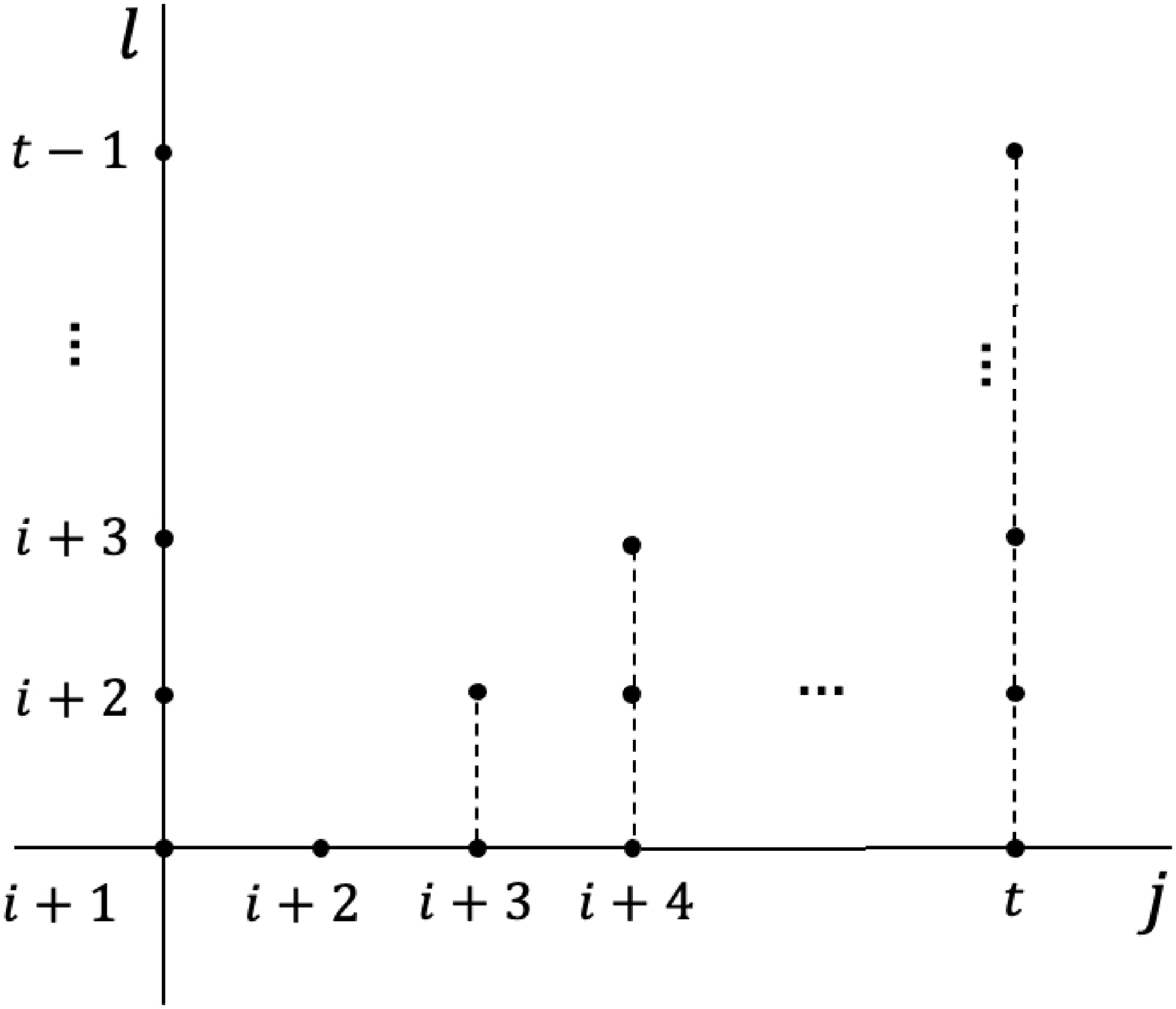}
\caption{Index variables used for the double-summation term in the \textrm{\textbf{Proof of Claim 3}} .}
\label{fig:notationchange}
\end{figure}
Coefficients for $y_i,i=1,...,t-2$ are
\ifCLASSOPTIONonecolumn
\begin{align*}
\tilde{F}_{i,t+1}=&B_{i,t+1}-\sum_{j=i+1}^tF_{i,j}B_{j,t+1}\\
=&B_{i,t+1}-[B_{i,i+1}B_{i+1,t+1}+\sum_{j=i+2}^t(B_{i,j}-\sum_{l=i+1}^{j-1}B_{i,l}F_{l,j})B_{j,t+1}]\\
=&B_{i,t+1}-(B_{i,i+1}B_{i+1,t+1}+\sum_{j=i+2}^tB_{i,j}B_{j,t+1}-\sum_{j=i+2}^t\sum_{l=i+1}^{j-1}B_{i,l}F_{l,j}B_{j,t+1}).\\
\end{align*}
\else
\begin{align*}
\tilde{F}_{i,t+1}=&B_{i,t+1}-\sum_{j=i+1}^tF_{i,j}B_{j,t+1}\\
=&B_{i,t+1}-[B_{i,i+1}B_{i+1,t+1}\\
&+\sum_{j=i+2}^t(B_{i,j}-\sum_{l=i+1}^{j-1}B_{i,l}F_{l,j})B_{j,t+1}]\\
=&B_{i,t+1}-(B_{i,i+1}B_{i+1,t+1}+\sum_{j=i+2}^tB_{i,j}B_{j,t+1}\\
&-\sum_{j=i+2}^t\sum_{l=i+1}^{j-1}B_{i,l}F_{l,j}B_{j,t+1}).\\
\end{align*}
\fi
With the order change of indices for the double summation term in the last equation, (see Fig. \ref{fig:notationchange} for reference) we will get
\ifCLASSOPTIONonecolumn
\begin{align*}
\tilde{F}_{i,t+1}=&B_{i,t+1}-(\sum_{j=i+1}^{t-1}B_{i,j}B_{j,t+1}+B_{i,t}B_{t,t+1}-\sum_{l=i+1}^{t-1}\sum_{j=l+1}^tB_{i,l}F_{l,j}B_{j,t+1})\\
=&B_{i,t+1}-[B_{i,t}B_{t,t+1}+\sum_{j=i+1}^{t-1}B_{i,j}(B_{j,t+1}-\sum_{l=j+1}^tF_{j,l}B_{l,t+1})]\\
=&B_{i,t+1}-(B_{i,t}F_{t,t+1}+\sum_{j=i+1}^{t-1}B_{i,j}F_{j,t+1})=F_{i,t+1}\\
\end{align*}
\else
\begin{align*}
\tilde{F}_{i,t+1}=&B_{i,t+1}-(\sum_{j=i+1}^{t-1}B_{i,j}B_{j,t+1}+B_{i,t}B_{t,t+1}\\
&-\sum_{l=i+1}^{t-1}\sum_{j=l+1}^tB_{i,l}F_{l,j}B_{j,t+1})\\
=&B_{i,t+1}-[B_{i,t}B_{t,t+1}\\
&+\sum_{j=i+1}^{t-1}B_{i,j}(B_{j,t+1}-\sum_{l=j+1}^tF_{j,l}B_{l,t+1})]\\
=&B_{i,t+1}-(B_{i,t}F_{t,t+1}+\sum_{j=i+1}^{t-1}B_{i,j}F_{j,t+1})=F_{i,t+1}\\
\end{align*}
\fi
where
\begin{equation*}
F_{i,t+1}=
	\begin{cases}
	B_{i,t+1}, & \text{for $i=t$,}\\
	B_{i,t+1}-\sum_{j=i+1}^tB_{i,j}F_{j,t+1}, & \text{for $1\leq i< t$}.
	\end{cases}
\end{equation*}
Note that we are considering the training phase, $t$ is upper-bounded by $T$ where $T$ denotes the training duration.
$\hfill\blacksquare$\\

\textbf{Proof of Claim 4}. Using $F_{i,j}$'s definition, its expectation is
\begin{align*}
\E[F_{i,j}]&=\E[B_{i,j}-\sum_{k=1}^{j-i-1}B_{i,j-k}F_{j-k,j}]\\
&=\E[B_{i,j}]-\sum_{k=1}^{j-i-1}\E[B_{i,j-k}F_{j-k,j}]
\end{align*}
when $1\leq i<j\leq T+1$ where $T$ is the training duration. Note that $B_{i,j-k}$ and $F_{j-k,j}$ are uncorrelated when $1\leq k\leq j-i-1$ and the distribution of $B_{i,j},1\leq i<j\leq T+1$ is independent of its indices $i$ and $j$ when $\textbf{a}_n$ is considered following the same distribution for $1\leq n\leq T+1$. Denote $\E[B_{i,j}]=b$, we get
\begin{equation*}
\E[F_{i,j}]=b\left(1-\sum_{k=1}^{j-i-1}\E[F_{j-k,j}]\right).
\end{equation*}
It can be verified that
\begin{equation}
\label{equ:ebfb}
\E[F_{j-1,j}]=\E[{B_{j-1,j}}]=b
\end{equation}
which justifies the claimed equality for $i=j-1,1<j\leq T+1$. Suppose there exists $1\leq t\leq T-1$ such that the claimed equality holds for $i=j-l,l<j\leq T+1$ when $1\leq l\leq t$, then for $i=j-(t+1),t+1<j\leq T+1$ we have
\begin{align*}
\E[F_{j-t-1,j}]&=b\left(1-\sum_{k=1}^t\E[F_{j-k,j}]\right)\\
&=b\left(1-\frac{b[1-(1-b)^t]}{1-(1-b)}\right)=b(1-b)^t
\end{align*}
Then, substituting $t=T$, we get the expectation of the sum of $F_{i,T+1}$ with all qualified $i$, i.e., $1\leq i\leq T$, as
\begin{align*}
    \E\left[\sum_{i=1}^TF_{i,T+1}\right]=\sum_{i=1}^T\E[F_{i,T+1}]
    =&\sum_{i=1}^Tb(1-b)^{T-i}\\
    =&1-(1-b)^T.
\end{align*}
$\hfill\blacksquare$\\

\bibliographystyle{IEEEtran}
\bibliography{ref}

\newpage
\begin{IEEEbiography}[{\includegraphics[width=1in,height=1.25in,clip,keepaspectratio]{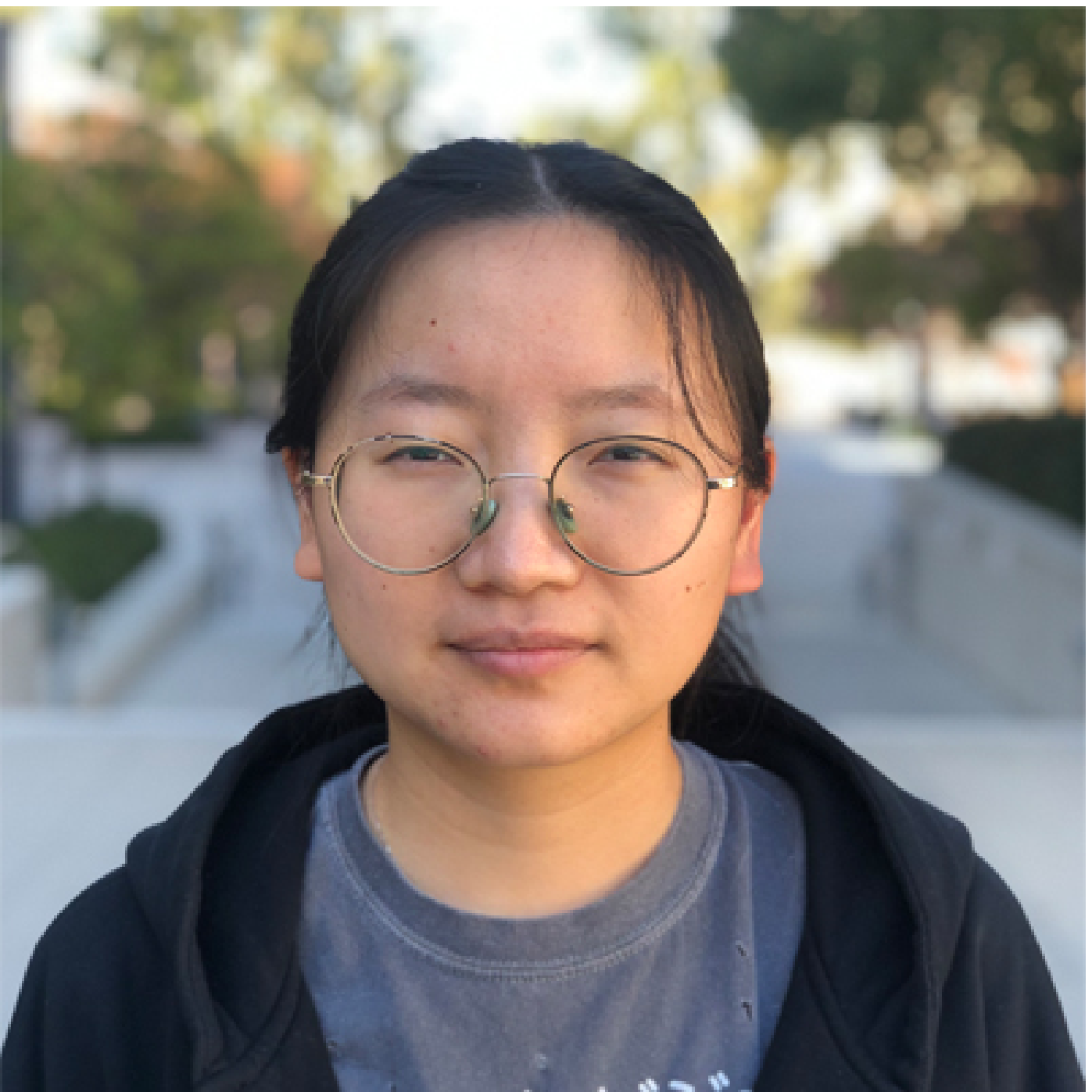}}]{Yue Zhao} is currently a Ph.D. candidate at the Department of Electrical Engineering and Computer Science, University of California at Irvine, Irvine, USA. She got her bachelor's degree from Beijing Institute of Technology, Beijing, China in 2016 and her master's degree from University of California at Irvine, Irvine, USA in 2019. Her research interests mainly include communication systems and graph signal processing.
\end{IEEEbiography}
\begin{IEEEbiography}[{\includegraphics[width=1in,height=1.25in,clip,keepaspectratio]{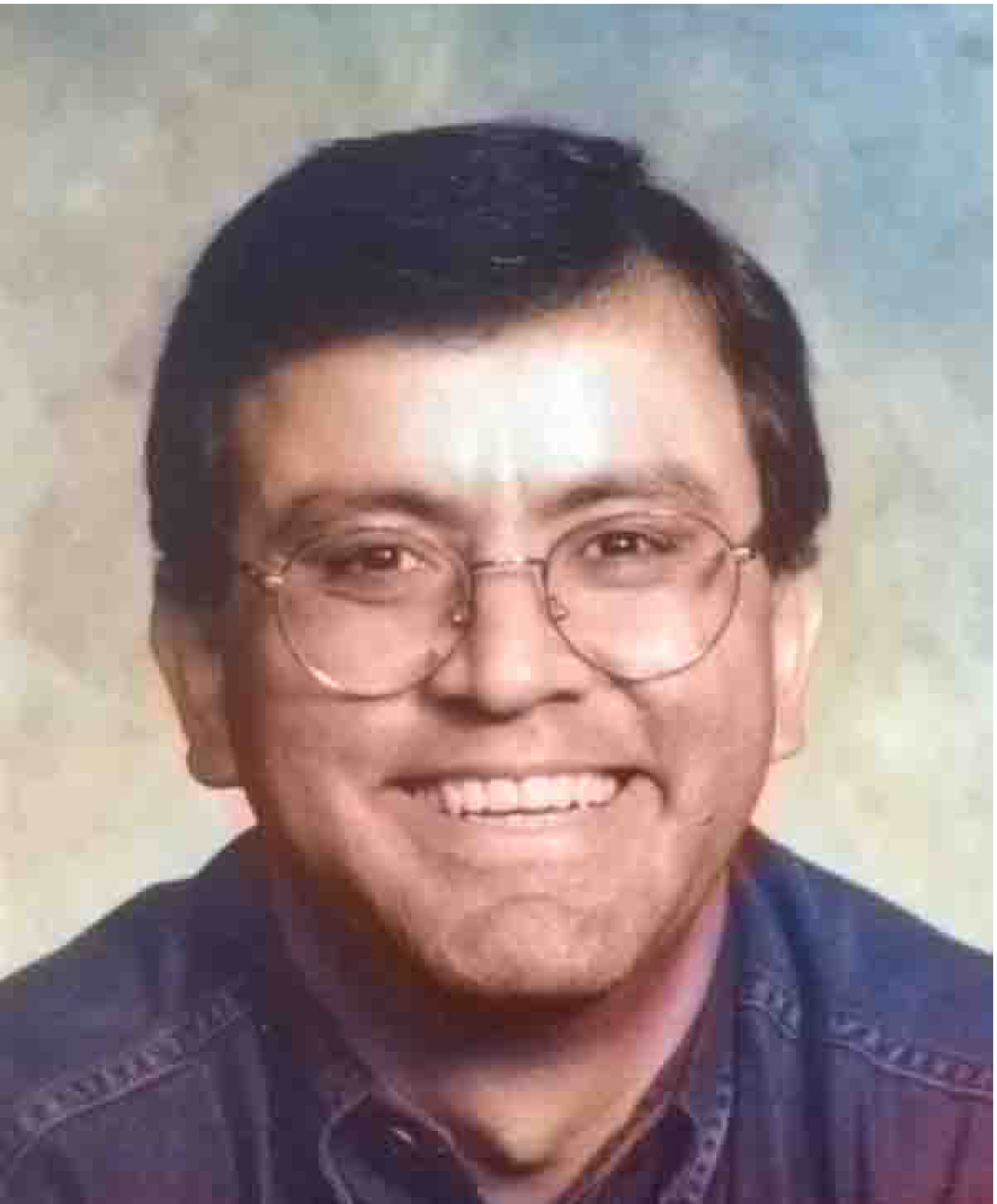}}]{Ender Ayanoglu} received the Ph.D. degree in electrical engineering from Stanford University, Stanford, CA, USA, in 1986. He was with the Bell Labs Communications Systems Research Laboratory, Holmdel, NJ, USA. From 1999 to 2002, he was a Systems Architect with Cisco Systems, Inc., San Jose, CA, USA. Since 2002, he has been a Professor with the Department of Electrical Engineering and Computer Science, University of California at Irvine, Irvine, CA, USA, where he served as the Director of the Center for Pervasive Communications and Computing and the Conexant-Broadcom Endowed Chair from 2002 to 2010. He was a recipient of the IEEE Communications Society Stephen O. Rice Prize Paper Award in 1995 and the IEEE Communications Society Best Tutorial Paper Award in 1997. He received the IEEE Communications Society Communication Theory Technical Committee Outstanding Service Award in 2014. From 2000 to 2001, he served as the Founding Chair of the IEEE-ISTO Broadband Wireless Internet Forum, an industry standards organization. He served on the Executive Committee of the IEEE Communications Society Communication Theory Committee from 1990 to 2002 and as its Chair from 1999 to 2002. From 1993 to 2014, he was an Editor of the \textit{IEEE Transactions on Communications}. He served as the Editor-in-Chief for the \textit{IEEE Transactions on Communications} from 2004 to 2008 and the \textit{IEEE Journal on Selected Areas in Communications}-Series on Green Communications and Networking from 2014 to 2016. He was the Founding Editor-in-Chief of the \textit{IEEE Transactions on Green Communications and Networking} from 2016 to 2020. He is an IEEE Communications Society Distinguished Lecturer from 2022 to 2023. He has been an IEEE Fellow since 1998.
\end{IEEEbiography}

\vfill

\end{document}